\documentclass[sigconf]{acmart}

\usepackage{multirow}
\usepackage[normalem]{ulem}
\useunder{\uline}{\ul}{}
\usepackage{graphicx}
\usepackage{tabularx}
\usepackage{makecell}
\usepackage{array} 

\usepackage{longtable}

\usepackage{booktabs}

\usepackage{multirow}
\usepackage{makecell}
\usepackage{verbatim}
\usepackage{threeparttable}
\usepackage{lscape}
\copyrightyear{2025}
\acmYear{2025}
\setcopyright{cc}
\setcctype{by-nc-nd}
\acmConference[ICHEC 2025]{Proceedings of the 2025 International Conference on Human-Engaged Computing}{November 21--23, 2025}{Singapore, Singapore}
\acmBooktitle{Proceedings of the 2025 International Conference on Human-Engaged Computing (ICHEC 2025), November 21--23, 2025, Singapore, Singapore}
\acmDOI{10.1145/3786995.3787027}
\acmISBN{979-8-4007-2234-9/2025/11}

\begin{document}

\title[Evaluating Text-based Conversational Agents for Mental Health]{Evaluating Text-based Conversational Agents for Mental Health: A Systematic Review of Metrics, Methods and Usage Contexts}

\author{Jiangtao Gong}
\orcid{0000-0002-4310-1894}
\authornote{Corresponding Author}
\affiliation{%
  \institution{Institute of AI Research, Tsinghua University}
  \city{Beijing}
  \country{China}}
\email{gongjiangtao2@gmail.com}

\author{Xiao Wen}
\email{xiaowen@ncepu.edu.cn}
\orcid{0009-0009-5401-4209}
\affiliation{%
  \institution{Institute of AI Research, Tsinghua University}
  \city{Beijing}
  \country{China}
}

\author{Fengyi Tao}
\orcid{0009-0001-2478-5098}
\affiliation{%
  \institution{Institute of AI Research, Tsinghua University}
  \city{Beijing}
  \country{China}}
\email{fengyi.tao1234@gmail.com}

\author{Xinqi Wang}
\orcid{0009-0004-8987-9986}
\affiliation{%
  \institution{Institute of AI Research, Tsinghua University}
  \city{Beijing}
  \country{China}}
\email{po3ppy3mar@gmail.com}

\author{Xixi Yang}
\orcid{0009-0003-9560-7218}
\affiliation{%
  \institution{Institute of AI Research, Tsinghua University}
  \city{Beijing}
  \country{China}}
\email{xixiyang@link.cuhk.edu.cn}

\author{Yangrong Tang}
\orcid{0009-0008-9714-4321}
\affiliation{%
  \institution{Institute of AI Research, Tsinghua University}
  \city{Beijing}
  \country{China}}
\email{tangxtong2022@gmail.com}

\renewcommand{\shortauthors}{Gong, et al.}

\begin{abstract}
Text-based conversational agents (CAs) are increasingly used in mental health, yet evaluation practices remain fragmented. We conducted a PRISMA-guided systematic review (May–June 2024) across ACM Digital Library, Scopus, and PsycINFO. From 613 records, 132 studies were included, with dual-coder extraction achieving substantial agreement ($ \kappa $ = 0.77–0.92). We synthesized evaluation approaches across three dimensions: metrics, methods, and usage contexts. Metrics were classified into CA-centric attributes (e.g., reliability, safety, empathy) and user-centric outcomes (experience, knowledge, psychological state, health behavior). Methods included automated analyses, standardized psychometric scales, and qualitative inquiry. Temporal designs ranged from momentary to follow-up assessments. Findings show reliance on Western-developed scales, limited cultural adaptation, predominance of small and short-term samples, and weak links between automated performance metrics and user well-being. We argue for methodological triangulation, temporal rigor, and equity in measurement. This review offers a structured foundation for reliable, safe, and user-centered evaluation of mental health CAs.
\end{abstract}





\maketitle

\section{Introduction}
The conversational agent (CA) is a conversational system, often referred to as a chatbot, that is capable of interacting with humans in natural language, such as text or voice~\cite{Kocaballi2019}. With the advancement of artificial intelligence, particularly the emergence of large language models, conversational agents have become pervasive in our daily lives~\cite{wu2024llm,bao2025generative,cao2025designing,zhang2025exploring,tang2025counselor}. In a world where mental health is increasingly prioritised, conversational agents designed for mental health support have attracted significant attention. Such applications are not limited to clinical contexts; CAs can be deployed in non-clinical settings through text, including stress intervention~\cite{Lin2023}, crisis care~\cite{Haque2023}, and psychological companionship~\cite{Ya-Hsin2023}. In the mental health domain, common examples of CAs include Woebot~\cite{Durden2023,Fitzpatrick2017}, Tess~\cite{Fulmer2018}, and ChatGPT~\cite{Franco2023}, which can provide 24-hour contact and companionship, as well as interventions based on cognitive behavioural therapy. These systems have been shown to exert positive effects on mental health, including reductions in depressive symptoms.

Therefore, it is not surprising that the application of CAs in the field of mental health has become a key topic of concern. The evaluation of such systems is intended to determine whether a CA is able to meet the needs of technical experts, mental health practitioners, and end users simultaneously. In practice, this is a considerable challenge. Firstly, a mental health CA should align with professional standards of psychological practice while also appealing to diverse user populations. However, mental health practitioners, who are important stakeholders, are often underrepresented in evaluation processes~\cite{Gooding2021}. Moreover, user experience and outcomes are complex, multidimensional, and context-dependent~\cite{Li2024722,Wilson2022}. Secondly, research is conducted across multiple disciplines—computer science, psychology, and medicine—each with distinct priorities. This often leads to evaluations that privilege one perspective, such as model performance or user satisfaction, at the expense of others. Consequently, evaluations may appear fragmented and are difficult to generalise across domains~\cite{Wilson2022}. It is therefore essential to provide a comprehensive overview of evaluation practices in order to inform more integrative design and application of CAs for mental health.

\begin{figure}
    \includegraphics[width=1.0\columnwidth]{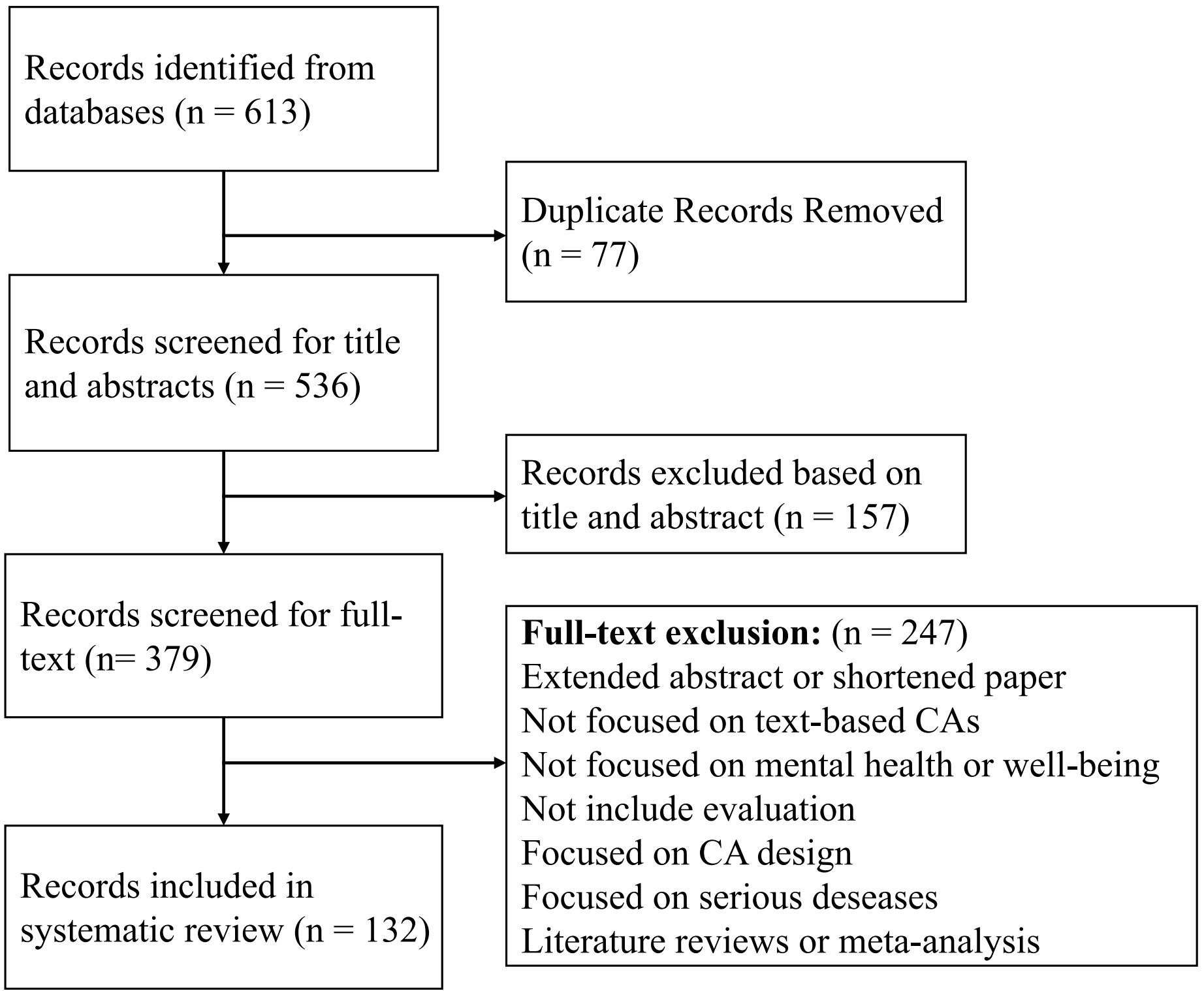}
    \caption{PRISMA flow diagram}
    \Description{Research method illustrated by the PRISMA flow diagram}
    \label{fig:prisma}    
\end{figure}

Prior reviews have examined CA assessment in specific domains, such as mental health~\cite{Jabir2023}, public health~\cite{Wilson2022}, and healthcare more broadly~\cite{AbbasianKhatibi-391}, often from either a technical or user-centred perspective. These reviews have highlighted the importance of defining clear metrics, yet their coverage remains partial. Some focused on functionality, usability, or engagement~\cite{XuejiaZhang2023}, while others emphasised the integration of behavioural change measures and usability indicators~\cite{Kocaballi2022}. Reviews of AI-supported healthcare conversations have provided useful taxonomies of user-centric metrics~\cite{AbbasianKhatibi-391}, but challenges persist in matching metrics with appropriate methods. Scoping reviews of CA interventions in mental health suggest the need for standardised measurement tools and stronger integration of technological and self-report approaches~\cite{Jabir2023}. Reviews from user experience and engagement perspectives further argue for systematic and multidimensional assessment frameworks~\cite{Feng2023,Khosravi2024}. Despite these contributions, gaps remain in linking CA-centric and user-centric evaluation, considering the temporal context of assessments, and identifying systemic limitations such as cultural bias or methodological fragmentation.

To address these challenges, we conducted a systematic review of prior research focused on the evaluation of CAs in mental health. Our objective is to synthesise existing evaluation practices across metrics, methods, contexts, and limitations. We therefore ask the following research questions:

\begin{enumerate}
    \item[RQ1:] What metrics have been used to evaluate text-based mental health CAs, and how can they be systematically organized into CA-centric attributes and user-centric outcomes?
    \item[RQ2:] Which methodological approaches are employed to assess these metrics, and how do different methods complement each other?
    \item[RQ3:] How are evaluation practices situated across different usage contexts, including the timing and duration of assessments?
    \item[RQ4:] What limitations in current evaluation practices emerge from the literature?
\end{enumerate}

To investigate these questions, we conducted a systematic review across major digital libraries in HCI, psychology, and medical informatics. From an initial pool of studies, we screened and retained 132 papers that met predefined inclusion criteria. All eligible works were coded for evaluation metrics, methodological approaches, contextual factors, and reported limitations, with iterative refinement to ensure consistency. The synthesized results are presented in a structured form that enables comparison across disciplines and highlights gaps in current practices. In sum, our contributions can be summarised as follows:
\begin{enumerate}
    \item We provide a comprehensive overview of evaluation metrics for mental health CAs, clarifying what aspects should be assessed.  
    \item We synthesise methodological approaches into a structured framework.  
    \item We examine the influence of evaluation contexts such as timing and duration on assessment design and validity.  
    \item We identify systemic limitations in current practices and propose directions toward more inclusive and rigorous evaluation frameworks.  
\end{enumerate}

\begin{table*}
\caption{Numbers of Study and Key characteristics of Samples}
\begin{threeparttable}
\begin{tabular}{|l|l|l|lllll|}
\hline
\multicolumn{1}{|c|}{\textbf{}} & \multicolumn{1}{c|}{\textbf{}} & \multicolumn{2}{c|}{\textbf{Sample Size}} & \multicolumn{4}{c|}{\textbf{Sample Characteristic}}                                                                                                                                                     \\ \hline
\textbf{Method}                 & \textbf{\begin{tabular}[c]{@{}l@{}}Number \\ of Studies\end{tabular}}      & \textbf{Mean}                             & \multicolumn{1}{l|}{\textbf{{[}Min, Max{]}}} & \multicolumn{1}{l|}{\textbf{Age}} & \multicolumn{1}{l|}{\textbf{{[}Min, Max{]}}} & \multicolumn{1}{l|}{\textbf{\begin{tabular}[c]{@{}l@{}}Including \\ Experts\tnote{1}\end{tabular}}} & \textbf{\begin{tabular}[c]{@{}l@{}}Including \\ Users\tnote{2}\end{tabular}} \\ \hline
Overall                         & 132                            & 465.06                                    & \multicolumn{1}{l|}{{[}2,36070{]}}           & \multicolumn{1}{l|}{28.12}        & \multicolumn{1}{l|}{{[}11,82{]}}        & \multicolumn{1}{l|}{16}                         & 126                      \\ \hline
Quantitative Research           & 126                            & 532.98                                    & \multicolumn{1}{l|}{{[}2,36070{]}}           & \multicolumn{1}{l|}{28.80}        & \multicolumn{1}{l|}{{[}11,82{]}}        & \multicolumn{1}{l|}{14}                         & 122                      \\ \hline
Qualitative Research           & 43                             & 79.68                                     & \multicolumn{1}{l|}{{[}3,766{]}}             & \multicolumn{1}{l|}{26.70}        & \multicolumn{1}{l|}{{[}14,80{]}}        & \multicolumn{1}{l|}{11}                         & 38                       \\ \hline
\end{tabular}
\begin{tablenotes}
        \footnotesize
        \item[1,2] Values below present the number of studies including experts or users as participants. Mixed methods studies are counted in both.
    \end{tablenotes}
\end{threeparttable}
\label{tab:overview}
\end{table*}

\section{Method}
Our review includes the following steps: 1) developing the search strategy and performing the literature search, 2) conducting the title and abstract screening, 3) completing a full-text screening, 4) running a quality assessment, and 5) extracting and analyzing the data. 
\subsection{Search Strategy}
According to the Preferred Reporting Items for Systematic Reviews and Meta-Analyses (PRISMA)~\cite{MoherShamseer-391}, our database search was conducted from May to June 2024. Fig.~ref{fig:prisma} provides an overview of our searching process and related search results. Searches of three databases identified 613 unique citations, including ACM Digital Library, Scopus and PsychInfo. These databases were chosen because they are well-known digital libraries within the fields of human computer interaction and psychology.

We used the Boolean “OR” to combine alternate terms (case insensitive) within each scope regarding mental health, evaluation as well as conversational agents, used “AND” to join the major related concepts and used "not" to exclude interference elimination concepts of congenital and degenerative diseases. Our search terms varied in form according to different databases, but the main content remained the same. Specifically, we apply the following search string to the metadata (including the title, keywords, and abstract) of the papers in each database (taking ACM Digital Library as an example, others showing in the Appendix):

\textit{("mental health" OR "mental wellness" OR wellbeing OR "well-being" OR "SWB" OR happiness OR happy OR "positive affect*" OR "negative affect*" OR "positive emotion*" OR "negative emotion*" OR mood OR "life satisfaction" OR "satisfaction with life") AND ("social bot*" OR "dialogue system*" OR "conversational agent*" OR "conversational bot*" OR "conversational system*" OR "conversational interface*" OR "chatbot*" OR "chat bot*" OR "chatterbot*" OR "chatter bot*" OR "chat-bot*" OR "smartbot*" OR "smart bot*" OR "smart-bot*" OR "digital assistant*" OR "counseling agent*" ) AND ("evaluation" OR "scale" OR "questionnaire" OR "guideline" OR "Review" OR "qualit*") AND NOT("autis*" OR "Alzheimer" OR "Parkinson" OR "disease" OR "*ache" OR "ADHD")}

\subsection{Screening}

We screened all search outcomes using a two-step process: a title-abstract review and a full-text review. Since 77 articles were duplicated, we removed the duplicated articles from the archive for this study. In the first step, the two researchers screened the title and abstract of each paper for its relevance to the topic of the study. If the title and abstract contain conversational agent, mental health, and assessment, the literature is retained. When inconsistencies and uncertainties arise, the two researchers discuss them. In this step, we excluded 157 papers and then reviewed the remaining 379 papers in a full-text screening.

In the second step, four researchers from the research team worked together to determine the final inclusion papers based on the constructed inclusion and exclusion criteria. Studies that satisfied all of the following inclusion criteria were included:  1) the paper is based on empirical research, and 2) the paper evaluates conversational agents for mental health.  Studies that met at least one of the following exclusion criteria were removed: 1) the article was not written in English, 2) the article was not peer-reviewed, 3) the article was an abstract,  or an extended abstract, 4) the article was secondary research, such as the review or meta-analysis,  5) The research population of the paper are patients with presence of somatic illness, neuro-developmental disorders, severe mental illness, or substance use disorder, who need professional medical assistance rather than CA's help. This criteria was chosen to make this review focus on non-clinical mental conditions~\cite{HERBENER2024100401}. 6) The paper focuses on CA's speech and virtual image, excluding text. 7) This article is an exploration of CA design principles and does not contain an evaluation of CA. The research team held regular meetings (e.g., weekly meetings and day-to-day online communication) to validate the inclusion and exclusion of articles.

\subsection{Quality Assessment}
In this session, members of the research team assessed the quality of the included articles and whether the  evaluation methods and evaluation metrics could be extracted from the articles. Also, if the included study selectively reports on the results of the assessment, it will be deleted. During this process, members of the research team continue to discuss to reach a consensus. 
After all the above processes, our systematic review included 132 papers.

\subsection{Data Extraction and Analysis}
We extract key information related to the research question through unanimous discussion. We mainly recorded the following information: year of publication, publication, methodology and sample size, characteristics of the study population, evaluation metrics (primary and secondary metrics), methods proposed for evaluation metrics, definitions or operational definitions of evaluation metrics, and methods used for evaluation metrics. Two researchers separately extracted information from each paper and recorded it in a spreadsheet. Differences in data extraction are compared and discussed at any time to reach a consensus. Due to the large amount of data, 20 random papers were extracted for back-to-back coding by 4 researchers in the research team, and a high inter-rating reliability (IRR) was obtained, with Cohen kappa coefficient of different dimensions ranging from 0.77 to 0.92.

We used frequency statistics to describe the basic situation of the articles in the database, and used bottom-up theme analysis to summarize and sort out the evaluation metrics and methods.

\section{Results}


\subsection{Overview}

In total, 132 studies involving human participants were identified. Among these, 126 employed questionnaires or standardized scales, 43 employed interviews, and several combined multiple evaluation methods. Table~\ref{tab:overview} summarises the methodological distribution and sample profiles across these studies.

Quantitative studies using questionnaires tended to recruit relatively large samples, with an average size of 532.98 participants. However, the majority of such studies remained modest in scale: 64.96\% (87/126) included fewer than 100 participants, and only 2.92\% (4/126) exceeded 1,000 participants. By contrast, qualitative studies had a markedly smaller average sample size of 79.68 participants. Within these, 18.60\% (8/43) involved more than 100 participants, while 83.72\% (36/43) reported below-average sample sizes, and a small proportion (9.30\%, 4/43) had fewer than 10 participants.

Across both methodological groups, there was limited variation in the reported age range of participants. As text-based CAs presuppose adequate literacy, no studies included children younger than 11 years, and the oldest reported participants were 82 years of age.

Taken together, this overview highlights the general methodological landscape of existing evaluations. The following sections address our research questions in detail, examining the specific metrics adopted (RQ1), the methods applied to assess them (RQ2), and the contexts in which these evaluations are situated (RQ3).

\subsection{Metrics of Evaluation (RQ1)}

This section outlines the taxonomy of metrics used to evaluate mental health CAs. To address what should be evaluated, we distinguish between two entities: the agent and the user. CA-centric metrics capture the inherent attributes and competence of the agent, while user-centric metrics reflect how people experience and are affected by the interaction. The latter are further divided into user experience and user changes. Table~\ref{tab:Metrics} provides an overview of sub-metrics and associated references, and the following subsections elaborate on each theme in detail.

\subsubsection{Agent attribute}

CA-centric evaluation focuses on the agent’s inherent properties, ranging from basic technical performance to domain-specific expertise. 

\textbf{Framework performance} covers foundational aspects such as response efficiency and resource usage, ensuring smooth and timely dialogue~\cite{8925455,10455044,s24030845,blagec2022global,Weeks2023}.  

\textbf{Algorithm performance} reflects the CA’s interaction capacity. Accuracy is commonly reported using precision and recall~\cite{Cheng2023535,Qiu2020381,Deepika20201196,Ngũnjiri2023,Bang2020230}, while similarity and semantic evaluation rely on BLEU, ROUGE-L, or perplexity (PPL)~\cite{Mishra2023,Zhu20233148,Goel2021}. Information quality is judged through text clarity and coherence~\cite{Dharrao20241,Beredo2022300,Saha2021,Farrand2024}, diversity is often quantified with Distinct-n indices~\cite{Mishra2023,Zhu20233148}, and conciseness reflects the CA’s ability to communicate succinctly~\cite{Gabrielli2020,Ly201739,Cheng2024140}.  

\textbf{Human-likeness} assesses the extent to which a CA simulates human characteristics. Emotional intelligence captures the recognition and expression of emotions~\cite{Elyoseph2023,Mansoori2022,Zhang2011412}, social intelligence refers to the ability to sustain relationships through contextually appropriate interaction~\cite{Hayashi2024521,Hopman2024161,Liao2023,Bravo2020941}, and personality addresses consistency of style and persona~\cite{Ahmad2022923,Schwartz2022139,Escobar-Viera2023}.  

\textbf{Reliability} is critical for mental health applications. It encompasses safety, fairness, privacy, trustworthiness, explainability, and transparency. These dimensions ensure ethical operation, protect user data, reduce bias, and sustain user trust~\cite{Khosravi2024,Durden2023,Valtolina2021,Fadhil2018,Weeks2023,sarkar2023review,Ly201739}.  

Finally, \textbf{well-being expertise} evaluates the CA’s capacity to apply psychological knowledge, such as using therapeutic techniques (e.g., CBT, DBT)~\cite{Woodnutt202479,Kaywan2023,Hakani2022796} and supporting professionals through empowerment in their work~\cite{Blease2024}.  

Together, these dimensions provide a structured framework for assessing CA attributes in mental health contexts, from baseline performance to advanced clinical competence.

\subsubsection{User Experience}

User experience captures users’ subjective perceptions and behaviors after interacting with a CA, serving as a bridge between technical development and actual adoption. Sustained use and therapeutic effectiveness depend on positive user experiences, which can be grouped into three areas: usability, relationship, and individual engagement.  

\textbf{Usability} refers to ease of use, satisfaction, and effectiveness in meeting user goals. It is typically measured with scales such as the System Usability Scale (SUS) and Post-Study System Usability Questionnaire (PSSUQ)~\cite{Gabrielli2020,Cheng2024140}. Sub-dimensions include acceptability (general evaluation and willingness to use)~\cite{Valtolina2021,Li2024722,Wrightson-Hester2023}, ease of use (simplicity, accessibility, and functional availability)~\cite{Solanki2023143,Escobar-Viera2023,Nicol2022}, satisfaction (overall appraisal and continued use intention)~\cite{Fitzpatrick2017,Provoost2020,Liao2021}, and perceived agent performance (fluency, informativeness, innovativeness)~\cite{Mishra2023,Andrews2023418,Yang2019}.  

\textbf{Relationship} assesses the bonds formed between users and CAs. General human–AI relationships include self-disclosure, companionship, perceived closeness, and attachment~\cite{Chou2024,Christoforakos2021,Lisetti2013}. In the mental health context, therapeutic relationships are especially relevant, often measured with working alliance or client–counselor analogues to capture trust, collaboration, and adherence~\cite{Hopman2024161,Darcy2021,Provoost2020,Liu2022}.

\textbf{Individual engagement} reflects the depth of user participation. Dialogue retention quantifies conversation length and duration~\cite{Horii2019190,Mansoori2022,Mishra2023}, behavioral engagement captures participation and adherence~\cite{Kettle2021791,Liao2021,Moore2024}, cognitive engagement relates to mental effort and attentiveness~\cite{Gabrielli2021,Moilanen2022138,Williams2021}, and emotional engagement assesses affective responses such as mood change, often linked to therapeutic impact~\cite{Fitzpatrick2017,Valtolina2021,Bravo2020941}.  

Together, these metrics provide a multidimensional picture of user experience, connecting usability and relational factors with deeper engagement patterns, and setting the stage for assessing user-level changes.

\subsubsection{User Changes}

User change captures the psychological, behavioral, and social transformations that occur after interacting with a CA. As the ultimate goal of mental health CAs, these metrics reflect effectiveness and practical value in fostering well-being. They can be grouped into acquired knowledge, mental state, and health-related behaviors.  

\textbf{Acquired knowledge} refers to users’ improved mastery of psychological and health-related knowledge. This includes health information, such as mental health literacy and coping strategies~\cite{Ngũnjiri2023,Mishra2023,Cheng2024140}, as well as broader social and collaboration skills that help users navigate daily life and interpersonal contexts~\cite{Ngũnjiri2023,Karkosz2024}.  

\textbf{Mental state} measures changes in users’ psychological conditions, typically with validated scales or qualitative reports. Four aspects are commonly considered. First, overall well-being is assessed through life satisfaction and positive affect scales, with studies consistently showing improvements after CA interventions~\cite{Ehrlich2024435,Potts2023,Terblanche202220,Lin2023264,Ly201739,Liao2021,Maharjan2022,Williams2021,Suganuma2018,Sia202134,Lee2023,Hungerbuehler2021,Karkosz2024,Inkster2023}. Second, negative status improvement captures reductions in depression, anxiety, stress, and related states, one of the most widely reported outcomes across trials~\cite{Karkosz2022424,Chou2024,Lopatovska2022192,Rodríguez-Martínez2024,Durden2023,Daley2020,Terblanche202220,Kettle2021791,Ly201739,Allan2023,Farrand2024,Williams2021,Gabrielli2021,Romanovskyi2021ElomiaCT,Fitzpatrick2017,Wrightson-Hester2023,Shidara2024,Schroeder2018,Peuters2024,Li2024722,Darcy2021,Oliveira2021,Mauriello2021,Bendig2021,Nicol2022,He2022,Eagle2022,Hungerbuehler2021,Suharwardy2023,Inkster2023,Delahunty2018327,Kaywan2023,Suganuma2018,Schick2022,Escobar-Viera2023,Qiu2020381,Ehrlich2024435,Drouin2022,Fulmer2018,Ghandeharioun20198,Yang2019,Torkamaan2023,Sayis2024945,Provoost2020,Kurashige2019962}. Third, psychological competence highlights gains in resilience, self-efficacy, and cognitive flexibility~\cite{Lee2023,Terblanche202220,Peuters2024,Ellis-Brush2021187,Durden2023,Moore2024,Ehrlich2024435,Nicol2022,Schroeder2018,Wrightson-Hester2023,Lee2019,Gabrielli2021,Inkster2023,Oliveira2021,Shidara2024,Allan2023,Bendig2021,Benke2020,Ngũnjiri2023,Ta2020}. Finally, physiological metrics, such as heart rate or stress biomarkers, provide real-time, objective indicators that complement self-reports~\cite{Ciechanowski2019539}.  

\textbf{Health-related behavior} evaluates observable lifestyle and social changes. This includes the application of therapeutic approaches in daily life~\cite{Schroeder2018,Ta2020,Dosovitsky2021}, direct behavioral improvements such as reduced anxious behaviors or better sleep~\cite{Valtolina2021,Peuters2024,Hungerbuehler2021}, greater engagement in meaningful activities like exercise, work, or learning~\cite{Ta2020,Rodríguez-Martínez2024,Kettle2021791,Allan2023,Moore2024}, and enhanced social relationships that signal improved social functioning~\cite{Christoforakos2021,Chou2024,Peuters2024,Rodríguez-Martínez2024,Lopatovska2022192}.

\onecolumn
\begin{longtable}{|p{1.5cm}|p{1.5cm}|p{4cm}|p{9cm}|}
\caption{Evaluation Metrics of Conversational Agents in Mental Health\label{tab:Metrics}}\\
\hline
Theme & Sub Theme & Basic Metrics & Examples \\ 
\hline
\endhead

\hline
\endfoot

\multirow{19}{1.5cm}{Agent Attribute}  
  & \multirow{2}{1.5cm}{Framework Performance}    
  & Response Performance               
  & \cite{8925455,10455044,s24030845} \\ \cline{3-4}

  &                       
  & Resources Occupation               
  & \cite{blagec2022global,Weeks2023} \\ \cline{2-4}

  & \multirow{6}{1.5cm}{Algorithms Performance}   
  & Accuracy                           
  & \cite{Cheng2023535,Qiu2020381,10455044,Deepika20201196,Ngũnjiri2023,Bang2020230,Srivastava20231118,Mishra2023,Moore2024,s24030845,Moilanen2023,Lee2019} \\ \cline{3-4}

  &                          
  & Similarity and Semantic Evaluation 
  & \cite{Cheng2023535,Mishra2023,s24030845,Zhu20233148,Goel2021,Srivastava20231118,Rodríguez-Martínez2024,Ngũnjiri2023,Feng2023} \\ \cline{3-4}

  &                          
  & Information Quality                 
  & \cite{Dharrao20241,Cheng2023535,Mishra2023,Beredo2022300,Saha2021,Crasto2021,Bang2020230,Lubis2018876,Brännström2024,Weeks2023,Farrand2024,Williams2021,Ngũnjiri2023,Woodnutt202479,Khosravi2024,Garcia-Velez2024389} \\ \cline{3-4}

  &                          
  & Diversity                          
  & \cite{Mishra2023,Zhu20233148,Bang2020230,Cheng2023535} \\ \cline{3-4}

  &                          
  & Conciseness                        
  & \cite{Mishra2023,Ngũnjiri2023,Gabrielli2020,Ahmad2022923,Ly201739,Cheng2024140,Karkosz2024} \\ \cline{2-4}

  & \multirow{3}{1.5cm}{Human-likeness}  
  & Emotional Intelligence             
  & \cite{Cheng2023535,Elyoseph2023,Mansoori2022,s24030845,Zhang2011412} \\ \cline{3-4}

  &                          
  & Social Intelligence                
  & \cite{Hayashi2024521,Mishra2023,Ngũnjiri2023,Moore2024,s24030845,Hopman2024161,Torkamaan2023,Brännström2024,Jabir2024,Liao2023,Maharjan2021,Bravo2020941,Lisetti2013} \\ \cline{3-4}

  &                          
  & Personality                        
  & \cite{Hopman2024161,Ahmad2022923,Karkosz2024,Schwartz2022139,Valtolina2021,Cameron2019121,Escobar-Viera2023,Weeks2023,Boyd2022,Drouin2022,Cheng2023535,Qiu2020381} \\ \cline{2-4}

  & \multirow{6}{1.5cm}{Reliability}  
  & Safety                             
  & \cite{Khosravi2024,Rodríguez-Martínez2024,Moilanen2023,sarkar2023review,Valtolina2021,Niess2018916,Ahmad2022923,Durden2023} \\ \cline{3-4}

  &                          
  & Fairness                           
  & \cite{Weeks2023} \\ \cline{3-4}

  &                          
  & Privacy                            
  & \cite{Weeks2023,Khosravi2024,Andrade-Arenas2024114,Ahmad2021,Benke2020,Fadhil2018,Li2024722,Ngũnjiri2023,Bendig2021} \\ \cline{3-4}

  &                          
  & Trustworthiness                     
  & \cite{Valtolina2021,Horii2019190,Fadhil2018,Moilanen2023,Liao2023,Ahmad2022923,Weeks2023} \\ \cline{3-4}

  &                          
  & Explainability                     
  & \cite{sarkar2023review} \\ \cline{3-4}

  &                          
  & Transparency                       
  & \cite{Ahmad2022923,Ly201739} \\ \cline{2-4}

  & \multirow{2}{1.5cm}{Expertise}      
  & Therapy Technique                  
  & \cite{Woodnutt202479,Ngũnjiri2023,Kaywan2023,Hakani2022796,Kaywan2021193} \\ \cline{3-4}

  &                          
  & Work Empowerment                   
  & \cite{Blease2024} \\ \hline

\multirow{2}{1.5cm}{User Experience}  
  & Usability                
  & Acceptability                      
  & \cite{Valtolina2021,Bendig2021,Wrightson-Hester2023,Rodríguez-Martínez2024,Li2024722,Nicol2022,Karkosz2024,Bravo2020941,Moore2024} \\ \cline{3-4}

  &                          
  & Ease of Use                        
  & \cite{Solanki2023143,Weeks2023,Schwartz2022139,Cheng2024140,Boian2024,Boyd2022,Escobar-Viera2023,He2022,Williams2021,Lisetti2013,Gabrielli2020,Bravo2020941,Qiu2020381,Provoost2020,Chou2024,Shah20221229,Barbosa2022355,Rodríguez-Martínez2024,Woodnutt202479,Ahmad2022923,Kaywan2023,Nicol2022,Ngũnjiri2023,Moore2024,Khosravi2024,Andrews2023418,Wrightson-Hester2023} \\ \cline{3-4}

  &                          
  & Satisfaction                       
  & \cite{Wrightson-Hester2023,Fitzpatrick2017,Schwartz2022139,Escobar-Viera2023,Anmella2023,Provoost2020,Andrade-Arenas2024114,Gabrielli2020,Suharwardy2023,Pangsrisomboon202246,Williams2021,Fulmer2018,Cheng2024140,Kaywan2023,Shah20221229,Kettle2021791,Benke2020,Hopman2024161,Weeks2023,Valtolina2021,Ngũnjiri2023,Liao2021,Dosovitsky2021,Pérez-Marín2013955,Andrews2023418,Bravo2020941} \\ \cline{3-4}

  &                          
  & Perceived Agent Performance        
  & \cite{Mishra2023,Maharjan2021,s24030845,Ngũnjiri2023,Andrews2023418,Wrightson-Hester2023,Yang2019,Anmella2023,Gabrielli2020} \\ \cline{2-4}

  & \multirow{2}{1.5cm}{Relationship}  
  & General Human-AI Relationship      
  & \cite{Chou2024,Christoforakos2021,Ly201739,Lee2019,Sayis2024945,Liao2021,Fadhil2018,Andrews2023418,Peuters2024,Kuhlmeier202230,Niess2018916,Lisetti2013} \\ \cline{3-4}

  &                          
  & Therapeutic Relationship            
  & \cite{Hopman2024161,Hopman2023,Karkosz2024,Suharwardy2023,Darcy2021,Chou2024,Ellis-Brush2021187,He2022,Andrews2023418,Peuters2024,Niess2018916,Ahmad2022923,Ngũnjiri2023,Provoost2020,Williams2021,Liu2022} \\ \cline{2-4}

  & \multirow{4}{1.5cm}{Individual Engagement}    
  & Dialogue Retention                 
  & \cite{Horii2019190,Peuters2024,Christoforakos2021,Mansoori2022,Mishra2023,Liao2021,8925455} \\ \cline{3-4}

  &                          
  & Behavioral Engagement              
  & \cite{Liao2021,Kettle2021791,Provoost2020,Peuters2024,Karkosz2024,Christoforakos2021,Wrightson-Hester2023,Moore2024,8925455,Maharjan2022,Qiu2020381} \\ \cline{3-4}

  &                          
  & Cognitive Engagement               
  & \cite{Qiu2020381,Moilanen2022138,Gabrielli2021,Gkinko20221714,Williams2021,Moore2024} \\ \cline{3-4}

  &                          
  & Emotional Engagement               
  & \cite{Fadhil2018,Valtolina2021,Andrews2023418,Fitzpatrick2017,Bravo2020941,Gkinko20221714,Maharjan2021} \\ \hline

\multirow{10}{1.5cm}{User Changes}     
  & \multirow{2}{1.5cm}{Acquired Knowledge}       
  & Health Information                 
  & \cite{Ngũnjiri2023,Mishra2023,Cheng2024140} \\ \cline{3-4}

  &                          
  & Social and Collaboration Skill     
  & \cite{Ngũnjiri2023,Karkosz2024} \\ \cline{2-4}

  & \multirow{4}{1.5cm}{Mental State}             
  & Overall Well-being                  
  & \cite{Ehrlich2024435,Potts2023,Terblanche202220,Lin2023264,Ly201739,Liao2021,Maharjan2022,Williams2021,Suganuma2018,Sia202134,Lee2023,Hungerbuehler2021,Karkosz2024,Inkster2023}  \\ \cline{3-4}

  &                          
  & Negative Status Improvement        
  & \cite{Karkosz2022424,Chou2024,Lopatovska2022192,Rodríguez-Martínez2024,Durden2023,Daley2020,Terblanche202220,Kettle2021791,Ly201739,Allan2023,Farrand2024,Williams2021,Gabrielli2021,Romanovskyi2021ElomiaCT,Fitzpatrick2017,Wrightson-Hester2023,Shidara2024,Schroeder2018,Peuters2024,Li2024722,Darcy2021,Oliveira2021,Mauriello2021,Bendig2021,Nicol2022,He2022,Eagle2022,Hungerbuehler2021,Suharwardy2023,Inkster2023,Delahunty2018327,Kaywan2023,Suganuma2018,Schick2022,Escobar-Viera2023,Qiu2020381,Ehrlich2024435,Drouin2022,Fulmer2018,Ghandeharioun20198,Yang2019,Torkamaan2023,Sayis2024945,Provoost2020,Kurashige2019962} \\ \cline{3-4}

  &                          
  & Psychological Competence           
  & \cite{Lee2023,Terblanche202220,Peuters2024,Ellis-Brush2021187,Durden2023,Moore2024,Ehrlich2024435,Nicol2022,Schroeder2018,Wrightson-Hester2023,Lee2019,Gabrielli2021,Inkster2023,Oliveira2021,Shidara2024,Allan2023,Bendig2021,Benke2020,Ngũnjiri2023,Ta2020} \\ \cline{3-4}

  &                          
  & Physiological metrics           
  & \cite{Ciechanowski2019539} \\ \cline{2-4}

  & \multirow{4}{1.5cm}{Health-related Behaviors} 
  & Therapeutic Approaches Application 
  & \cite{Schroeder2018,Ta2020,Dosovitsky2021} \\ \cline{3-4}

  &                          
  & Direct Behaviors Change            
  & \cite{Valtolina2021,Peuters2024,Hungerbuehler2021} \\ \cline{3-4}

  &                          
  & Daily Activities Engagement        
  & \cite{Ta2020,Rodríguez-Martínez2024,Kettle2021791,Allan2023,Moore2024} \\ \cline{3-4}

  &                          
  & Social Relationship                
  & \cite{Christoforakos2021,Chou2024,Peuters2024,Rodríguez-Martínez2024,Lopatovska2022192} \\ \hline

\end{longtable}
\twocolumn

Together, these metrics demonstrate that CAs not only provide immediate psychological support but also foster long-term improvements in knowledge, mental states, and behaviors, ultimately contributing to sustainable mental well-being.

\subsection{Methods of Evaluation (RQ2)}

Building on the taxonomy of metrics outlined in RQ1, this section examines how these metrics are operationalised in practice. We identified 47 studies employing automatic or semi-automatic approaches, supplemented by traditional user-centered methods. Evaluation methods can be broadly distinguished by their \textit{data sources}—ranging from system parameters, interaction logs, to physiological monitoring~\cite{Ciechanowski2019539,Peuters2024}—and by their \textit{analytical techniques}, including general statistical measures, custom advanced indicators, and NLP-specific metrics~\cite{s24030845,Cheng2023535,Mishra2023}. Together, these approaches illustrate the methodological diversity in assessing both CA attributes and user outcomes.  

The following subsections elaborate on three main categories of methods, highlighting their applications and limitations in the context of mental health CAs.

\subsubsection{Automatic Evaluation} 

Among the final set of literature, 33 papers involved automatic evaluation, including 4 review articles. Based on the metrics discussed in these reviews, an additional 18 relevant papers were identified, resulting in 47 references in total.

Automatic evaluation metrics for text-based mental health CAs span beyond framework and algorithm performance to cover user experience aspects, such as safety, reliability, and even mental health–related signals like physiological responses~\cite{Ciechanowski2019539} and emotional change~\cite{s24030845,Kurashige2019962,1909.06670,9030346}. We group these methods by \textit{statistical approach} and \textit{data source}, as summarized in Table~\ref{tab:Auto}.

\textbf{General statistical metrics} are widely used to assess efficiency and interaction quality. From \textit{system-level data}, metrics such as \textbf{CPU runtime}, \textbf{bug logs}, and \textbf{error rates} evaluate stability and performance during operation~\cite{Boian2024,Anmella2023,Lee2019}. From \textit{user interaction logs}, metrics like the \textbf{average number of turns}, \textbf{task completion rate}, \textbf{session length}, and \textbf{response latency} capture engagement and usability~\cite{Mishra2023,Farrand2024,Peuters2024,8925455}. Word count features, such as words per response, are also frequently used to measure depth of engagement~\cite{Liao2023,Lee2019}.

\textbf{Custom advanced metrics} extend these basic measures with domain-specific algorithms. Examples include \textbf{diagnostic accuracy}, \textbf{recommendation accuracy}, and \textbf{match-rate}, reflecting the CA’s ability to provide clinically appropriate outputs~\cite{Wang2020378,wei2018task,Deepika20201196}. Dialogue quality is captured through metrics such as \textbf{emotional expression intensity}~\cite{Feng2023} and \textbf{politeness scores}~\cite{Mishra2023}, while safety is assessed using tools like \textbf{lexicons} and resources such as ProsocialDialog and DiSafety~\cite{sarkar2023review,2305.08010,2205.12688,meade2023using}.

\textbf{NLP-specific metrics} remain central to automatic evaluation. Accuracy and fluency of generated text are measured using \textbf{BLEU}~\cite{s24030845,Cheng2023535,Mishra2023,papineni2002bleu}, \textbf{ROUGE}~\cite{lin2004rouge,Srivastava20231118}, \textbf{METEOR}~\cite{banerjee2005meteor}, and \textbf{BERTScore}~\cite{1904.09675}. Diversity is assessed with \textbf{Distinct-n} and its variants~\cite{Zhu20233148,Feng2023,Bang2020230}, while language fluency and uncertainty are captured through \textbf{perplexity (PPL)}~\cite{Beredo2022300}. Advanced models like SimCSE~\cite{Cheng2023535}, MoverScore~\cite{zhao2019moverscore}, and NUBIA~\cite{kane2020nubia} add semantic-level evaluation.

Finally, \textbf{physiological and behavioral monitoring} provides objective complements to self-reports. Wearable devices track \textbf{physical activity and sleep patterns}~\cite{Peuters2024}, while psychophysiological signals such as \textbf{ECG, EDA, and EMG} capture users’ emotional states during CA interaction~\cite{Ciechanowski2019539}. Sentiment and emotion analysis tools further quantify \textbf{affective changes} across interactions~\cite{9030346,1909.06670}.

Together, these methods illustrate the breadth of automatic evaluation, from low-level system efficiency to nuanced assessments of human–AI interaction quality.

\begin{table*}
\caption{Automatic Evaluation Metrics and Methods}
\label{tab:Auto}
\begin{tabular}{|>{\raggedright\arraybackslash}p{2cm}|>{\raggedright\arraybackslash}p{5cm}|>{\raggedright\arraybackslash}p{4cm}|>{\raggedright\arraybackslash}p{5cm}|}
\hline
{\color[HTML]{1F2329} \textbf{Data Source | Statistical Method}}&
  {\color[HTML]{1F2329} \textbf{  General Statistical Metrics}}&
  {\color[HTML]{1F2329} \textbf{  Custom Advanced Metrics}}&
  {\color[HTML]{1F2329} \textbf{  
NLP-Specific Metrics}}\\ \hline
{\color[HTML]{1F2329} \textbf{ Algorithm/ System Design Parameters; Resource Usage Data; Algorithm Output Results}}&
  {\color[HTML]{1F2329} Average number of turns~\cite{Mishra2023}; 
Average response length~\cite{s24030845};  Bug logs~\cite{Anmella2023};  
CPU Runtime~\cite{Boian2024}; 
Error rate~\cite{Lee2019};   Task completion rate~\cite{Mishra2023};  Tolerance of calls ~\cite{Anmella2023}}&
  {\color[HTML]{1F2329} Average politeness score~\cite{Mishra2023}; 
Controllability of response generation~\cite{Bang2020230}; 
Diagnostic accuracy~\cite{Wang2020378,10.1609/aaai.v33i01.33017346,wei2018task}; 
Emotional expression intensity~\cite{Feng2023};  Intent/story accuracy~\cite{Deepika20201196}; 
Match-rate~\cite{10.5555/3327757.3327834,8461918,10.1609/aaai.v33i01.33017346,wei2018task,2305.12474};  
Safety~\cite{Durden2023,2305.08010,2205.12688,meade2023using} }&
  {\color[HTML]{1F2329} {\color[HTML]{1F2329} BERT Score~\cite{ Srivastava20231118,1904.09675}; 
BLEU~\cite{s24030845,Zhu20233148,Cheng2023535,Feng2023,Mishra2023,Saha2021,Goel2021,papineni2002bleu};  Cosine similarity~\cite{Cheng2023535};  
Distinct~\cite{Zhu20233148,Cheng2023535,Feng2023,Bang2020230}; 
Meteor~\cite{Srivastava20231118,Mishra2023,banerjee2005meteor}; 
MoverScore~\cite{zhao2019moverscore}; 
NIST Score~\cite{Mishra2023,blagec2022global}; 
Perplexity~\cite{Dharrao20241,Cheng2023535,Feng2023,Mishra2023,Beredo2022300,Saha2021,Crasto2021,Bang2020230,Lubis2018876}; 
ROUGE~\cite{Srivastava20231118,Cheng2023535,Mishra2023,lin2004rouge};  Semantic relations~\cite{kane2020nubia}; 
TER~\cite{blagec2022global}}}\\ \hline
{\color[HTML]{1F2329} \textbf{User-System Interaction Data}}& {\color[HTML]{1F2329} 
In-app duration~\cite{Peuters2024}; 
Adherence~\cite{Provoost2020}; 
Number of interactions~\cite{Farrand2024,Fitzpatrick2017,8461918,10.1609/aaai.v33i01.33017346,wei2018task}; 
User engagement~\cite{Provoost2020,Farrand2024}; 
User recruitment and retention~\cite{Ehrlich2024435}; 
User response latency and frequency~\cite{8925455}; 
Word count~\cite{Liao2023,Lee2019,1909.06670}}& {\color[HTML]{1F2329} Emotional consistency~\cite{s24030845}; 
Feeling polarity change~\cite{Kurashige2019962}; 
User emotion~\cite{1909.06670}; 
User sentiment~\cite{9030346}}& {\color[HTML]{1F2329} User utterance~\cite{tang2025counselor}}\\ \hline
{\color[HTML]{1F2329} \textbf{Automated Monitoring Data During User Interaction}}&
  {\color[HTML]{1F2329} Electrocardiography~\cite{Ciechanowski2019539}; 
Electrodermal activity~\cite{Ciechanowski2019539}; 
Electromyography~\cite{Ciechanowski2019539};  Physical (in)activity and sleep~\cite{Peuters2024}; }&
  {\color[HTML]{1F2329} } &
  {\color[HTML]{1F2329} } \\ \hline
\end{tabular}

\end{table*}

\subsubsection{Standardized Scales}

Standardized scales constitute one of the most widely adopted tools for evaluating conversational agents in mental health research. They enable consistent, comparable, and psychometrically validated measurement of outcomes, ranging from user experience to mental health indicators~\cite{Fitzpatrick2017,Provoost2020,Nicol2022}.

Of the final set of literature, 70 papers employed standardized scales, including six review articles. From these reviews, 18 additional studies were identified, resulting in 81 papers in total. Across these studies, 125 distinct scales were initially reported, of which 109 were retained after excluding 12 with unclear citations and 4 reported only in German. All retained scales had documented reliability and validity.

\textbf{Geographic distribution.} The majority of scales originated in Western contexts, particularly the United States (n=67) and Europe (n=30), followed by Australia (n=6), Canada (n=6), and New Zealand (n=2). A smaller number were developed in Asia (Singapore n=2, China n=1, Japan n=1). Two scales were globally validated, and five were developed across multiple regions, reflecting broader cross-cultural applicability.

\textbf{Scale length.} Most instruments were concise: 54 contained ten or fewer items, 35 included 11–20 items, 11 ranged from 21–30 items, and 4 had 31–40 items. Only 7 exceeded 40 items. Some scales had multiple validated versions with different item lengths, resulting in overlaps across categories. This prevalence of brief instruments aligns with practical needs in CA studies, where participant burden must be minimized~\cite{Andrews2023418,Wrightson-Hester2023}.

\textbf{Temporal framing.} Of the 109 scales, 45 specified explicit time frames for assessment, enabling non-momentary evaluations such as mental health over the past week or month. These included coverage of today (n=6), yesterday (n=2), past week (n=15), past two weeks (n=9), past month (n=7), and longer intervals (six months to one year). In contrast, 80 scales lacked explicit time framing. Among these, 42 were used for momentary evaluations (e.g., immediate user experience or usability of CAs~\cite{Schroeder2018,Liao2021}), while 38 measured broader constructs such as life satisfaction or resilience. Five scales were applicable across multiple time frames, demonstrating their flexibility for both ecological momentary assessment and longitudinal designs~\cite{Hopman2024161,Peuters2024}.

In sum, standardized scales provide a robust foundation for CA evaluation, ensuring methodological rigor while accommodating diverse temporal and contextual requirements.

\subsubsection{Qualitative Research}

Qualitative methods provide indispensable insights into user experiences, perceptions, and contextual factors that cannot be captured by quantitative measures alone~\cite{Provoost2020,Hopman2024161,Rodríguez-Martínez2024}. They are especially valuable in mental health research, where subjective experiences and interpersonal dynamics strongly influence outcomes. To ensure generalizability, this review excluded studies limited to specific interventions, product names, or time-bound events, while retaining methods that broadly reflect the evaluation of conversational agents (CAs).

Eight main qualitative approaches were identified (Table~\ref{tab:qual}).  

\textbf{Interviews} are the most widely used method, offering in-depth exploration of user experiences, needs, and acceptance of CAs~\cite{Wrightson-Hester2023,Peuters2024}. While user interviews capture perceptions of usability, engagement, or trust~\cite{Benke2020,Rodríguez-Martínez2024}, expert interviews emphasize clinical validity and technical robustness~\cite{Eagle2022,Kuhlmeier202230}. Frameworks such as UTAUT and ACTS are frequently employed to guide structured questioning~\cite{Valtolina2021,Hopman2024161}.

\textbf{Focus groups} extend this approach by leveraging group dynamics to reveal collective perspectives and stimulate discussion~\cite{Rodríguez-Martínez2024,Ngũnjiri2023}. They are particularly useful in adolescent and community settings, providing insights into feasibility and acceptability of CAs among young or vulnerable populations~\cite{Wrightson-Hester2023}.

\textbf{Think-aloud protocols} capture real-time cognitive processes as users verbalize their thoughts while interacting with CAs, highlighting usability barriers and decision-making patterns~\cite{mauriello2021suite}. Similarly, \textbf{card-sorting} tasks assess how users categorize system features, offering evidence on information architecture and navigation preferences~\cite{mauriello2021suite}.

\textbf{Diary studies} allow longitudinal tracking of user experiences, documenting day-to-day interactions and the evolving role of CAs in users’ routines~\cite{Lopatovska2022192}.  

\textbf{Open-ended survey questions} balance scalability with qualitative depth, capturing perceptions of trust, emotional impact, and therapeutic expectations without restricting participants to fixed responses~\cite{Ahmad2021,Fitzpatrick2017,Elyoseph2023}.  

\textbf{Product review analysis} draws on naturally occurring user feedback in app stores or online platforms to assess adoption, usability, and concerns such as privacy or anthropomorphism~\cite{haque2022app,prakash2020intelligent}.  

\textbf{Conversational analysis} examines chat logs to evaluate dialogue quality, engagement levels, and therapeutic responsiveness~\cite{Lee2019,Inkster2023,martinengo2022evaluation}.  

Finally, \textbf{desk research} provides contextual understanding by synthesizing organizational reports, project documents, and grey literature~\cite{Gkinko20221714,gkinko2021ai}.  

Together, these qualitative approaches complement automatic metrics and standardized scales by foregrounding the lived experiences of users. They not only identify usability and relational dynamics but also uncover nuanced barriers and facilitators that shape long-term CA adoption and mental health impact.

In sum, three complementary approaches are used to evaluate conversational agents in mental health. \textbf{Automatic metrics} offer efficiency and reproducibility, providing insights into system performance, language quality, and user engagement at scale. \textbf{Standardized scales} supply reliable and validated instruments to assess mental health outcomes and user experiences, ensuring comparability across studies. \textbf{Qualitative methods} add depth by foregrounding user voices, uncovering nuanced perspectives, and contextualizing quantitative findings. Together, these methods form a multi-layered evaluation framework that balances objectivity, validity, and experiential richness. This comprehensive approach not only enables robust assessment of CA effectiveness but also highlights the trade-offs between efficiency, generalizability, and depth. Building on this foundation, the next section (RQ3) turns to the question of how evaluation designs are structured, examining their temporal scope, experimental settings, and alignment with real-world applications.

\subsection{Evaluation Contexts (RQ3)}

This subsection examines how evaluation practices of conversational agents (CAs) are situated across different usage contexts, with a particular focus on the role of time. Temporal design is a critical factor, as both the duration of interaction with a CA and the timing of assessment shape the reliability and validity of findings. While many studies administered evaluations immediately after an intervention, others employed longer intervention periods or delayed assessments, ranging from days to several months. The choice of temporal design reflects trade-offs between feasibility and the capacity to capture meaningful changes in mental health outcomes. In what follows, we first discuss the duration of CA-user interactions and then turn to the timing of outcome assessments.


\subsubsection{Interaction Duration}

A total of 19 articles specified user interaction or intervention periods in their methodology. Since mental health outcomes typically emerge through repeated engagement rather than single exposures, studies with longer durations or multiple measurement points provide stronger evidence. The most common intervention period was \textbf{2–4 weeks}, adopted for its balance of feasibility and effectiveness. Such short-term designs resemble brief counseling programs, often sufficient to detect initial improvements in mood or behavior.

In contrast, interventions lasting \textbf{less than one week} are generally limited to evaluating immediate responses such as emotional state or trust, rather than enduring mental health changes~\cite{Kraus2021}. Slightly longer designs of \textbf{5–8 weeks} were also common, with eight weeks serving as a practical threshold for longitudinal tracking. For example, several studies implemented repeated measures at 2, 4, 6, and 8 weeks to capture the trajectory of user engagement and psychological outcomes~\cite{Suharwardy2023,Ehrlich2024435,Durden2023}.

Extended periods of \textbf{12–14 weeks} align with the timeframes of standard intermediate counseling programs. Such designs, though resource-intensive, allow for more robust evaluation of sustained effects. For instance, the ChatPal study recruited participants across multiple European countries to test outcomes over a 14-week period~\cite{Potts2023}. Finally, a small number of studies conducted \textbf{six-month interventions}, aiming to track long-term behavior change and goal attainment, which is ambitious even by clinical counseling standards~\cite{TerblancheMolyn-391}.

Together, these findings show that interaction duration is not uniform but varies according to research objectives, ranging from short-term usability checks to long-term assessments of behavioral change. This diversity underscores the importance of aligning temporal design with the intended outcomes of evaluation.

\subsubsection{Evaluation Duration} 
Evaluation duration refers to the length of time participants spend completing an assessment. Methodological design in this area requires careful balance: overly long assessments risk participant fatigue and unreliable responses, while overly short assessments may not capture sufficient information for meaningful analysis.  

\textbf{Interviews and Focus Groups.} Interview duration varies according to study goals. Brief interviews of less than 15 minutes are suitable for gathering targeted feedback or immediate impressions~\cite{Benke2020,Liao2021}, while in-depth interviews lasting 30–60 minutes (and occasionally up to 2 hours) allow exploration of complex experiences and therapeutic processes~\cite{Hopman2024161,Eagle2022}. Focus groups are typically longer, averaging 90–120 minutes, as group interaction encourages participants to elaborate and exchange perspectives~\cite{Rodríguez-Martínez2024,Wrightson-Hester2023}.  

\textbf{Diary Studies.} Longitudinal diary methods provide another perspective on duration. Participants typically contribute short daily entries of less than 10 minutes, recorded across 5–15 days, thereby producing cumulative datasets that capture temporal fluctuations in engagement and emotional state~\cite{Lopatovska2022192}. Workplace studies sometimes adjust this design by excluding weekends to focus on weekday routines.  

\textbf{Standardized Scales.} As discussed in before, scale-based evaluations also differ in their temporal framing. Of the 109 retained scales, 45 specify fixed timeframes ranging from a single day to six months. By contrast, 80 scales omit explicit timeframes, allowing flexible application either as momentary assessments of system usability or as broader indicators of mental health outcomes such as resilience or life satisfaction. This flexibility enables researchers to adapt scale usage to both short-term interaction studies and long-term outcome tracking.

In summary, evaluation contexts for mental health conversational agents are shaped by two critical dimensions: \textit{interaction duration} and \textit{assessment duration}. Studies commonly adopt short-to-medium interaction periods of 2–8 weeks, balancing feasibility with the need to capture meaningful changes, though longer studies of 12–24 weeks are occasionally employed for more sustained interventions. In terms of assessment, interviews and focus groups typically range from 15 minutes to 2 hours, diary methods distribute short assessments over multiple days, and standardized scales exhibit varied timeframes from daily check-ins to multi-month follow-ups. Together, these designs highlight the methodological trade-offs between immediacy and depth, participant burden and data richness. For future research, more systematic justification of timing decisions and increased use of longitudinal and multi-point assessments would enhance both the validity and ecological relevance of evaluation practices.

\onecolumn
\begin{longtable}{|p{2cm}|p{5cm}|p{9cm}|}
\caption{Qualitative Evaluation Methods\label{tab:qual}}\\
\hline
\multicolumn{1}{|c|}{\textbf{Methods}} & 
\multicolumn{1}{c|}{\textbf{Definition}} & 
\multicolumn{1}{c|}{\textbf{Example}} \\ \hline
\endhead

\hline
\endfoot

Interview
& Interviews with users/experts, using semi-structured, unstructured, or structured formats, to gather detailed insights into experiences, opinions, and needs related to mental health agents. 
& Artificial Intelligence~\cite{Wrightson-Hester2023,Gkinko20221714,gkinko2021ai} Mobile Phone~\cite{Wrightson-Hester2023,mauriello2021suite}  Emotions, Emotions at Work, Digital Workplace~\cite{Gkinko20221714,gkinko2021ai}  Mental Health Promotion, Change Interventions, Self-Regulation Techniques, Universal Prevention, Mobile Health Applications~\cite{Peuters2024}  Digital Wellbeing, Emotional Health, Loneliness, Conversational Chatbot, Social Support~\cite{Rodríguez-Martínez2024} Technology Use, AI Use, Digital Transformation, Digital Ways of Working~\cite{Gkinko20221714} Chatbot Use, Technology Adoption, AI Adoption~\cite{gkinko2021ai} Stress Relief, Stress Management, Virtual Agent~\cite{mauriello2021suite} Health Anxiety, Anthropomorphism, User Experience~\cite{Goonesekera2022}  Emotion Management,Team Communication~\cite{Benke2020} Voice agents,  Healthcare~\cite{Eagle2022} Acceptability, Feasibility~\cite{Wrightson-Hester2023} Dialogflow  Messenger~\cite{Andrade-Arenas2024114} Usability~\cite{Valtolina2021} 
\\ \hline

Focus Group
& A moderated group discussion with users or experts to explore their collective views, experiences,  and attitudes toward mental health agents. 
& Chatbot~\cite{Rodríguez-Martínez2024,Ngũnjiri2023,rathnayaka2022mental} Conversational Agents, Artificial Intelligence~\cite{Wrightson-Hester2023,rathnayaka2022mental}  Adolescent~\cite{Ngũnjiri2023, Wrightson-Hester2023}  Feasibility, Acceptability, Mobile Phone, Young People, Artificial  Therapist, Child, Social Support, Support~\cite{Wrightson-Hester2023} Personalised Assistance, Emotional Support, Mental Health Suppor,  Behavioural Activation, Mental Health Monitoring~\cite{rathnayaka2022mental} Older Adults, Social Support, Digital Wellbeing, Emotional Health, Loneliness~\cite{Rodríguez-Martínez2024} 
\\ \hline

Think Aloud  
& Users verbalize their thoughts while interacting with mental health agents, providing insights into their thought processes and usability concerns.
& Virtual agent, Therapy, Stress relief, Stress management, Mental health, Support, Mobile phone~\cite{mauriello2021suite}
\\ \hline

Card-Sorting
& Users organize information or features related to mental health agents into categories to assess their understanding and preferences for navigation and structure.
& Virtual agent, Therapy, Stress relief, Stress management, Mental health, Support, Mobile phone~\cite{mauriello2021suite}
\\ \hline

Diary Study 
& Participants record their interactions with mental health agents over time, capturing detailed insights into their ongoing experiences and any issues encountered.
& Wellbeing intervention, Conversational agent, Voice user interface, Adolescence~\cite{Lopatovska2022192}
\\ \hline

Open-ended Question of Survey 
& Surveys with open-ended questions that gather qualitative feedback from users about their experiences and perceptions of mental health agents.
& Mental health~\cite{Moilanen2023,koulouri2022chatbots,Ahmad2021,Fulmer2018,Fitzpatrick2017,Gabrielli2020}  Artificial intelligence~\cite{Elyoseph2023,koulouri2022chatbots,Ta2020,Fulmer2018} Anxiety~\cite{Williams2021,Fulmer2018,Fitzpatrick2017}  Depression~\cite{Fulmer2018,Fitzpatrick2017}  Acceptance and commitment therapy~\cite{peltola2023developing,Bendig2021} EHealth, Health system navigation, Electronic health care~\cite{noble2022developing}  Emotion elicitation, Chat-based  dialogue system~\cite{lubis2019positive} User expectations, Psychological flexibility~\cite{peltola2023developing}  Software agent,Therapeutic writing~\cite{Bendig2021}  Psychotherapy~\cite{Elyoseph2023}  Emerging adults, Stress~\cite{Williams2021}  College students~\cite{Fitzpatrick2017}  User experience~\cite{Yang2019} User Compliance~\cite{Torkamaan2023}  Personality-Adaptive Conversational Agents~\cite{Ahmad2021}  Social isolation~\cite{Dosovitsky2021} Well-being intervention~\cite{Gabrielli2020}     
\\ \hline

Product Reviews Analysis
& Examination of user reviews and feedback about mental health agents to identify strengths, weaknesses, and overall user opinions.
& Mobile applications~\cite{haque2022app,ahmed2022thematic}  Mental health~\cite{haque2022app,prakash2020intelligent} Mental Health Chatbots, Technology Adoption,  Privacy Calculus, Anthropomorphism~\cite{prakash2020intelligent} Social support, Artificial agents, Interpersonalrelations~\cite{Ta2020} Anxiety, Depression~\cite{ahmed2022thematic} 
\\ \hline

Conversational Analysis
& Detailed examination of interactions between users and mental health agents to understand communication patterns, effectiveness, and areas for improvement.
& Maternal mental health and wellbeing, Mental health~\cite{Inkster2023,Lee2019} Conversational agent, Depression~\cite{Inkster2023,martinengo2022evaluation}  Mood disorders, mHealth, Digital health~\cite{martinengo2022evaluation} Artificial intelligence, Compassion~\cite{Lee2019} 
\\ \hline

Desk Research
& Review of existing literature, reports, and documentations on mental health agents to gather background information and contextual understanding.
& Future of work, Digital workplace, Emotions at work, Emotions, Artificial intelligence,Chatbot~\cite{Gkinko20221714,gkinko2021ai} Digital ways of working, Digital transformation, AI use, Technology use~\cite{Gkinko20221714} Technology adoption, AI adoption~\cite{gkinko2021ai}
\\ \hline

\end{longtable}
\twocolumn

\section{Discussion}
\subsection{Summary of Key Findings and Contributions}

This review systematically examined the evaluation practices of text-based conversational agents (CAs) in mental health, synthesizing evidence from 132 studies. Several major findings emerged across our four research questions. First, the \textbf{overview} revealed a growing reliance on human-centered evaluation, with 126 studies using questionnaires or scales and 43 employing interviews, while highlighting the predominance of small to medium sample sizes. Second, in addressing \textbf{RQ1 (Evaluation Metrics)}, we identified a comprehensive framework of user-centered changes—acquired knowledge, mental states, and health-related behaviors—as well as automatic evaluation metrics and standardized scales. This classification demonstrates that evaluations extend beyond algorithmic performance, encompassing psychosocial outcomes and real-world behavioral impacts. Third, for \textbf{RQ2 (Evaluation Methods)}, we reviewed the use of standardized scales and diverse qualitative approaches such as interviews, focus groups, diary studies, and conversational analyses, showing how these methods complement quantitative measures by capturing contextualized and process-oriented insights. Fourth, in \textbf{RQ3 (Evaluation Contexts)}, we found that temporal design is a critical but underexplored factor: most studies adopt short-to-medium interaction durations (2–8 weeks), while assessment duration varies widely across interviews, diary studies, and standardized scales, influencing both data validity and participant burden. Finally, in \textbf{RQ4 (Synthesis)}, we proposed an integrative framework connecting evaluation metrics, methods, and contexts, offering a holistic perspective for future research design.

Our study makes three distinct contributions. First, it \textbf{expands the scope of evaluation} by systematically combining automatic measures, standardized scales, and qualitative methods into a unified framework. This multi-dimensional approach reflects both technical robustness and human-centered well-being, bridging a gap identified in earlier reviews that treated these domains separately. Second, we \textbf{highlight the overlooked role of evaluation contexts}, particularly the timing of interaction and assessment, which prior literature rarely considered despite their methodological importance for validity and ecological relevance. Third, by synthesizing diverse metrics and methods across disciplines, this review \textbf{establishes a comprehensive taxonomy of evaluation practices}, providing researchers and practitioners with practical guidance for designing, selecting, and justifying evaluation strategies in future CA studies.

In sum, this review advances the field by shifting the evaluation of mental health CAs from fragmented, tool-specific practices toward a structured, integrative framework. By situating evaluation within the interplay of metrics, methods, and contexts, it enables more rigorous, user-centered, and comparable assessments, thereby supporting the development of conversational agents that are not only technically sound but also effective, safe, and impactful in promoting mental health.

\subsection{Limitations in Current Evaluation Practices (RQ4)}

This review identifies several limitations in the current evaluation practices for mental health conversational agents (CAs). These limitations reflect methodological constraints, reliance on Western-developed tools, the short-term and small-scale nature of many studies, and weak links between automated metrics and actual user outcomes.

\subsubsection{Temporal Facets of Studies}
A major limitation lies in the treatment of temporal factors in evaluation. Many studies do not clearly distinguish between users’ mental states (e.g., emotions, cognition) and behaviors, often mixing short-term self-reports with long-term behavioral outcomes. This raises concerns about whether effects observed in short-term trials can be sustained over time~\cite{Kraus2021}. In psychological research, multi-time-point designs often measure emotion first, followed by cognitive or behavioral outcomes, which helps establish causal pathways. However, most CA evaluations remain short-term, with interventions lasting less than 8 weeks~\cite{Durden2023,Suharwardy2023}. While this design minimizes participant attrition, it limits the ability to assess enduring improvements in mental health. In contrast, medical or clinical CA applications often require long-term engagement, such as interventions for substance use disorders, which extend beyond 8 weeks~\cite{Prochaska2021}. Thus, short study windows in non-clinical contexts may provide only partial evidence of CAs’ effectiveness.

\subsubsection{Weaknesses in Automated Evaluation}
Automated metrics offer efficiency and scalability but face significant limitations. Current approaches fall into three categories:  
(1) \textbf{General statistical metrics}, such as CPU runtime or error rates~\cite{Boian2024,Lee2019}, and log-file derived measures of engagement~\cite{Farrand2024,Fitzpatrick2017,Provoost2020};  
(2) \textbf{Custom advanced metrics}, such as politeness scores or polarity change~\cite{Mishra2023,Kurashige2019962};  
(3) \textbf{NLP-specific metrics}, such as perplexity and diversity~\cite{Dharrao20241,Cheng2023535,Feng2023}.  

Despite their variety, these metrics struggle to capture the nuanced emotional and psychological dimensions central to mental health interventions. For example, sentence-level emotional intensity can be quantified automatically, but holistic measures of therapeutic alliance still require human annotation~\cite{Feng2023}. Moreover, reliance on system-based data (e.g., configuration parameters) dominates current practice, while real-time psychophysiological data (e.g., EEG, ECG) remain rare due to high costs~\cite{Ciechanowski2019539}. With the rise of large language models (LLMs), some studies propose LLMs as evaluators, yet only a handful systematically explore this direction~\cite{AbbasianKhatibi-391,pan2024}, and most lack validation against expert judgment. This creates a weak link between automated evaluation and user outcomes.

\subsubsection{Dependence on Western-Developed Scales}
Standardized scales form a cornerstone of CA evaluation, but most are developed and validated in Western contexts. Of the 109 scales analyzed, the majority lack cultural adaptation, raising concerns about bias when applied in non-Western populations. While shorter scales are user-friendly, they risk omitting depth, whereas longer scales increase fatigue and dropout rates. Additionally, most scales are self-reported, making them vulnerable to social desirability bias. Although scales cover diverse timeframes (from daily to yearly recall), many lack flexibility to capture longitudinal change. These issues highlight the need for culturally inclusive, context-specific, and CA-oriented scale development.

\subsubsection{Neglect of Professional Competencies}
Finally, current practices rarely assess the professional or therapeutic abilities of CAs. Evaluations typically focus on user outcomes, such as symptom change, rather than whether the agent demonstrates core counseling skills like empathy or alliance-building. This is a significant gap compared to psychology, where competency models and 360-degree feedback are common~\cite{Allan1997}. Moreover, psychometric and diagnostic accuracy in CAs is often evaluated through algorithmic comparisons, but rigorous benchmarking against expert clinical judgment is limited~\cite{FanSun-391,SpeerPerrotta-392}. Without incorporating professional competency models and cross-validating psychometric outputs, CA evaluations risk overlooking key aspects of therapeutic quality.

In sum, limitations in current evaluation practices include:  
1) short-term and small-sample studies that cannot capture long-term outcomes;  
2) automated metrics that emphasize technical performance over therapeutic impact;  
3) reliance on Western-developed, self-report scales with limited cultural adaptation; and  
4) insufficient attention to professional competencies and diagnostic rigor.  
Addressing these gaps requires integrating longitudinal designs, developing culturally inclusive scales, combining automated and human evaluation, and establishing competency frameworks for CAs in mental health contexts.

\subsection{Future Research Directions in Wellbeing CA evaluation}

Future work should prioritize the development of more comprehensive and culturally adaptive evaluation frameworks for mental health CAs. First, establishing a \textbf{professional competency model} is essential. Drawing inspiration from 360-degree assessments in psychology and management~\cite{Allan1997}, researchers can design frameworks that capture core therapeutic skills such as empathy, alliance-building, and professional empowerment, rather than focusing solely on symptom reduction. This model should be complemented by structured progress indicators that track the development of CA competence over time.

Second, \textbf{advancing psychometric and diagnostic validation} is a critical next step. While current evaluations often compare algorithmic detection accuracy against computerized benchmarks, future studies should systematically align CA assessments with human expert judgments and clinical standards~\cite{FanSun-391,SpeerPerrotta-392}. This alignment will strengthen credibility in psychology and clinical practice. Moreover, mixed-method approaches—combining qualitative insights with scale-based assessments—can provide a more nuanced understanding of diagnostic validity.

Third, there is a pressing need to \textbf{design culturally inclusive and CA-specific scales}. Most standardized instruments remain Western-centric and self-reported, limiting their applicability across diverse populations. Future research should focus on adapting existing scales for non-Western contexts and developing new tools tailored to the unique interaction dynamics of CAs in mental health.

Finally, \textbf{integrating human and automated evaluations} will be key to advancing the field. Automated metrics, particularly those leveraging large language models, hold promise for scalable evaluation, but they must be validated against human expert assessment and linked directly to user outcomes. Future studies should explore hybrid frameworks that combine real-time automated monitoring with expert-driven evaluations to capture both technical performance and therapeutic quality.

In sum, future research must move beyond narrow or short-term evaluations and aim for comprehensive, culturally sensitive, and professionally grounded approaches. Such advances will enhance the reliability, validity, and clinical relevance of mental health CA assessments.

\subsection{Limitations of Research Methodology}

This review has several methodological limitations that should be acknowledged. First, despite systematic retrieval and screening, the inclusion of literature may still be biased toward English-language and peer-reviewed publications, potentially overlooking relevant studies in non-English contexts or unpublished gray literature. This could amplify the Western-centric perspective already present in many standardized evaluation tools. Second, the heterogeneity of study designs posed challenges in synthesizing findings. Studies varied widely in intervention duration, evaluation scales, and outcome reporting, which limited the possibility of direct comparisons and meta-analysis. Third, our classification of evaluation methods—into automated metrics, standardized scales, and qualitative approaches—was guided by the available literature. While this framework provides clarity, it inevitably simplifies overlapping practices and may obscure hybrid or emerging methodologies. Finally, although we aimed to capture temporal factors in evaluation (e.g., interaction and assessment duration), our analysis was constrained by the inconsistent and sometimes incomplete reporting of methodological details in primary studies. These limitations suggest that future reviews should adopt broader inclusion strategies, incorporate non-English literature, and apply mixed-method evidence synthesis to achieve a more comprehensive understanding of evaluation practices in mental health CAs.

\section{Conclusions}
Through our systematic review of 132 papers, this work paints a picture of the situation relevant to the assessment of CA in the field of mental health. We determined the whole process of this work, including the evaluation metrics, evaluation methods, and evaluation time. This paper calls attention to the complexities and differences that exist in this interdisciplinary discipline and identifies ways in which future research can address these major issues. Our research will help improve the reliability and validity of the evaluation process and results as researchers begin to normalize the use and reporting of information on important metrics and methods. CA research in the field of mental health, under the influence of the convergence of technological developments and psychology, is a challenging but crucial topic that deserves ongoing research.

\begin{acks}
This work is supported by National Natural Science Foundation Youth Fund 62202267, Beijing Natural Science Foundation L233033, and Beijing Municipal Science and Technology Project (Nos. Z231100010323005).
\end{acks}

\bibliographystyle{ACM-Reference-Format}
\bibliography{main}

@String{Computing = "Computing" }

@String{Computer = "{IEEE} Computer" }

@String{Springer = "Springer-Verlag" }

@INPROCEEDINGS{8925455,
  author={Ghandeharioun, Asma and McDuff, Daniel and Czerwinski, Mary and Rowan, Kael},
  booktitle={2019 8th International Conference on Affective Computing and Intelligent Interaction (ACII)}, 
  title={EMMA: An Emotion-Aware Wellbeing Chatbot}, 
  year={2019},
  volume={},
  number={},
  pages={1-7},
  doi={10.1109/ACII.2019.8925455}}

@INPROCEEDINGS{10455044,
  author={Limbachia, Janak and Damani, Yash and Dave, Shubh and Sagvekar, Vidya},
  booktitle={2023 6th International Conference on Advances in Science and Technology (ICAST)}, 
  title={MOODIFY: Tailored, Personal and Multifaceted AI Assistant for Young Adult Mental Health Issues}, 
  year={2023},
  volume={},
  number={},
  pages={106-110},
  doi={10.1109/ICAST59062.2023.10455044}}

@Article{s24030845,
AUTHOR = {Zhou, Tiehua and Yu, Zihan and Wang, Ling and Ryu, Keun Ho},
TITLE = {A Mood Semantic Awareness Model for Emotional Interactive Robots},
JOURNAL = {Sensors},
VOLUME = {24},
YEAR = {2024},
NUMBER = {3},
ARTICLE-NUMBER = {845},
URL = {https://www.mdpi.com/1424-8220/24/3/845},
PubMedID = {38339563},
ISSN = {1424-8220},
DOI = {10.3390/s24030845}
}

@article{AbbasianKhatibi-391,
   Author = {Abbasian, Mahyar and Khatibi, Elahe and Azimi, Iman and Oniani, David and Shakeri Hossein Abad, Zahra and Thieme, Alexander and Sriram, Ram and Yang, Zhongqi and Wang, Yanshan and Lin, Bryant and Gevaert, Olivier and Li, Li-Jia and Jain, Ramesh and Rahmani, Amir M.},
   Title = {Foundation metrics for evaluating effectiveness of healthcare conversations powered by generative AI},
   Journal = {npj Digital Medicine},
   Volume = {7},
   Number = {1},
   Pages = {82},
   DOI = {10.1038/s41746-024-01074-z},
   Year = {2024} }

@CONFERENCE{Cheng2023535,
	author = {Cheng, Jiale and Sabour, Sahand and Sun, Hao and Chen, Zhuang and Huang, Minlie},
	title = {PAL: Persona-Augmented Emotional Support Conversation Generation},
	year = {2023},
	journal = {Proceedings of the Annual Meeting of the Association for Computational Linguistics},
	pages = {535 – 554},
	url = {https://www.scopus.com/inward/record.uri?eid=2-s2.0-85175473067&partnerID=40&md5=82f2580a522e15826c129a383b4f3916},
	type = {Conference paper},
	publication_stage = {Final},
	source = {Scopus},
	note = {Cited by: 4},
        booktitle={}
}

@ARTICLE{Qiu2020381,
	author = {Qiu, Sihang and Gadiraju, Ujwal and Bozzon, Alessandro},
	title = {Just the right mood for HIT!: Analyzing the role of worker moods in conversational microtask crowdsourcing},
	year = {2020},
	journal = {Lecture Notes in Computer Science (including subseries Lecture Notes in Artificial Intelligence and Lecture Notes in Bioinformatics)},
	volume = {12128 LNCS},
	pages = {381 – 396},
	doi = {10.1007/978-3-030-50578-3_26},
	url = {https://www.scopus.com/inward/record.uri?eid=2-s2.0-85087041312&doi=10.1007%2f978-3-030-50578-3_26&partnerID=40&md5=21fe364c707912a33f42660901a0ff42},
	type = {Conference paper},
	publication_stage = {Final},
	source = {Scopus},
	note = {Cited by: 8; All Open Access, Green Open Access}
}

@CONFERENCE{Deepika20201196,
	author = {Deepika, Kanakamedala and Tilekya, Veeranki and Mamatha, Jatroth and Subetha, T.},
	title = {Jollity Chatbot- A contextual AI Assistant},
	year = {2020},
	journal = {Proceedings of the 3rd International Conference on Smart Systems and Inventive Technology, ICSSIT 2020},
	pages = {1196 – 1200},
	doi = {10.1109/ICSSIT48917.2020.9214076},
	url = {https://www.scopus.com/inward/record.uri?eid=2-s2.0-85094847216&doi=10.1109%2fICSSIT48917.2020.9214076&partnerID=40&md5=8511746c245c86a2fb24edb714f8579f},
	type = {Conference paper},
	publication_stage = {Final},
	source = {Scopus},
	note = {Cited by: 26}
}

@ARTICLE{Ngũnjiri2023,
	author = {Ngũnjiri, Anne and Memiah, Peter and Kimathi, Robert and Wagner, Fernando A. and Ikahu, Annrita and Omanga, Eunice and Kweyu, Emmanuel and Ngunu, Carol and Otiso, Lilian},
	title = {Utilizing User Preferences in Designing the AGILE (Accelerating Access to Gender-Based Violence Information and Services Leveraging on Technology Enhanced) Chatbot},
	year = {2023},
	journal = {International journal of environmental research and public health},
	volume = {20},
	number = {21},
	doi = {10.3390/ijerph20217018},
	url = {https://www.scopus.com/inward/record.uri?eid=2-s2.0-85176391770&doi=10.3390%2fijerph20217018&partnerID=40&md5=878e66a3bd92c7b02db89b675d5f52b1},
	type = {Article},
	publication_stage = {Final},
	source = {Scopus},
	note = {Cited by: 0; All Open Access, Gold Open Access, Green Open Access}
}

@article{sarkar2023review,
  title={A review of the explainability and safety of conversational agents for mental health to identify avenues for improvement},
  author={Sarkar, Surjodeep and Gaur, Manas and Chen, Lujie Karen and Garg, Muskan and Srivastava, Biplav},
  journal={Frontiers in Artificial Intelligence},
  volume={6},
  pages={1229805},
  year={2023},
  publisher={Frontiers Media SA}
}

@ARTICLE{Bang2020230,
	author = {Bang, Jeesoo and Han, Sangdo and Lee, Jong-Hyeok},
	title = {Listening-oriented response generation by exploiting user responses},
	year = {2020},
	journal = {Pattern Recognition Letters},
	volume = {140},
	pages = {230 – 237},
	doi = {10.1016/j.patrec.2020.10.007},
	url = {https://www.scopus.com/inward/record.uri?eid=2-s2.0-85093933794&doi=10.1016%2fj.patrec.2020.10.007&partnerID=40&md5=11ff72a30ca64c3d077514ebbbe67b16},
	type = {Article},
	publication_stage = {Final},
	source = {Scopus},
	note = {Cited by: 0}
}

@CONFERENCE{Delahunty2018327,
	author = {Delahunty, Fionn and Wood, Ian D. and Arcan, Mihael},
	title = {First insights on a passive major depressive disorder prediction system with incorporated conversational chatbot},
	year = {2018},
	journal = {CEUR Workshop Proceedings},
	volume = {2259},
	pages = {327 – 338},
	url = {https://www.scopus.com/inward/record.uri?eid=2-s2.0-85058207398&partnerID=40&md5=5cf206a2f3aa8d76a055d81df2de39a0},
	type = {Conference paper},
	publication_stage = {Final},
	source = {Scopus},
	note = {Cited by: 11}
}

@CONFERENCE{Srivastava20231118,
	author = {Srivastava, Aseem and Pandey, Ishan and Akhtar, Md Shad and Chakraborty, Tanmoy},
	title = {Response-act Guided Reinforced Dialogue Generation for Mental Health Counseling},
	year = {2023},
	journal = {ACM Web Conference 2023 - Proceedings of the World Wide Web Conference, WWW 2023},
	pages = {1118 – 1129},
	doi = {10.1145/3543507.3583380},
	url = {https://www.scopus.com/inward/record.uri?eid=2-s2.0-85159375944&doi=10.1145%2f3543507.3583380&partnerID=40&md5=b0bc3b0c102b2f67395fc114af8eefa3},
	type = {Conference paper},
	publication_stage = {Final},
	source = {Scopus},
	note = {Cited by: 3; All Open Access, Green Open Access}
}

@ARTICLE{Mishra2023,
	author = {Mishra, Kshitij and Firdaus, Mauajama and Ekbal, Asif},
	title = {GenPADS: Reinforcing politeness in an end-to-end dialogue system},
	year = {2023},
	journal = {PLoS ONE},
	volume = {18},
	number = {1 January},
	doi = {10.1371/journal.pone.0278323},
	url = {https://www.scopus.com/inward/record.uri?eid=2-s2.0-85145921838&doi=10.1371%2fjournal.pone.0278323&partnerID=40&md5=4f4608b8af2468b5ae9749d125bbca89},
	type = {Article},
	publication_stage = {Final},
	source = {Scopus},
	note = {Cited by: 5; All Open Access, Gold Open Access, Green Open Access}
}

@ARTICLE{Boian2024,
	author = {Boian, Rares and Bucur, Ana-Maria and Todea, Diana and Luca, Andreea and Rebedea, Traian and Podina, Ioana R.},
	title = {A conversational agent framework for mental health screening: design, implementation, and usability},
	year = {2024},
	journal = {Behaviour and Information Technology},
	doi = {10.1080/0144929X.2024.2332934},
	url = {https://www.scopus.com/inward/record.uri?eid=2-s2.0-85189348618&doi=10.1080%2f0144929X.2024.2332934&partnerID=40&md5=f0ecb76f42175d74ef86b7b35c878042},
	type = {Article},
	publication_stage = {Article in press},
	source = {Scopus},
	note = {Cited by: 0; All Open Access, Green Open Access}
}

@Article{Moore2024,
author="Moore, Richard
and Al-Tamimi, Abdel-Karim
and Freeman, Elizabeth",
title="Investigating the Potential of a Conversational Agent (Phyllis) to Support Adolescent Health and Overcome Barriers to Physical Activity: Co-Design Study",
journal="JMIR Form Res",
year="2024",
month="Jan",
day="31",
volume="8",
pages="e51571",
issn="2561-326X",
doi="10.2196/51571",
url="https://formative.jmir.org/2024/1/e51571",
url="https://doi.org/10.2196/51571",
url="http://www.ncbi.nlm.nih.gov/pubmed/38294857"
}

@ARTICLE{Moilanen2023,
	author = {Moilanen, Joonas and van Berkel, Niels and Visuri, Aku and Gadiraju, Ujwal and van der Maden, Willem and Hosio, Simo},
	title = {Supporting mental health self-care discovery through a chatbot},
	year = {2023},
	journal = {Frontiers in Digital Health},
	volume = {5},
	doi = {10.3389/fdgth.2023.1034724},
	url = {https://www.scopus.com/inward/record.uri?eid=2-s2.0-85150502421&doi=10.3389%2ffdgth.2023.1034724&partnerID=40&md5=9338f61a5fd0138a2dbef60b1318133b},
	type = {Article},
	publication_stage = {Final},
	source = {Scopus},
	note = {Cited by: 13; All Open Access, Gold Open Access, Green Open Access}
}

@inproceedings{Lee2019,
  title={Caring for Vincent: a chatbot for self-compassion},
  author={Lee, Minha and Ackermans, Sander and Van As, Nena and Chang, Hanwen and Lucas, Enzo and IJsselsteijn, Wijnand},
  booktitle={Proceedings of the 2019 CHI conference on human factors in computing systems},
  pages={1--13},
  year={2019}
}

@ARTICLE{Zhu20233148,
	author = {Zhu, Ying and Wang, Bo and Zhao, Dongming and Huang, Kun and Jiang, Zhuoxuan and He, Ruifang and Hou, Yuexian},
	title = {Grafting Fine-Tuning and Reinforcement Learning for Empathetic Emotion Elicitation in Dialog Generation},
	year = {2023},
	journal = {Frontiers in Artificial Intelligence and Applications},
	volume = {372},
	pages = {3148 – 3155},
	doi = {10.3233/FAIA230634},
	url = {https://www.scopus.com/inward/record.uri?eid=2-s2.0-85175849462&doi=10.3233%2fFAIA230634&partnerID=40&md5=be8f41515055f3a6653e827be00edb6e},
	type = {Conference paper},
	publication_stage = {Final},
	source = {Scopus},
	note = {Cited by: 2; All Open Access, Hybrid Gold Open Access}
}

@CONFERENCE{Goel2021,
	author = {Goel, Raman and Vashisht, Sachin and Dhanda, Armaan and Susan, Seba},
	title = {An Empathetic Conversational Agent with Attentional Mechanism},
	year = {2021},
	journal = {2021 International Conference on Computer Communication and Informatics, ICCCI 2021},
	doi = {10.1109/ICCCI50826.2021.9402337},
	url = {https://www.scopus.com/inward/record.uri?eid=2-s2.0-85104981204&doi=10.1109%2fICCCI50826.2021.9402337&partnerID=40&md5=10ab35052cab845ad082ba6bf13d2bfb},
	type = {Conference paper},
	publication_stage = {Final},
	source = {Scopus},
	note = {Cited by: 19}
}

@ARTICLE{Woodnutt202479,
	author = {Woodnutt, Samuel and Allen, Chris and Snowden, Jasmine and Flynn, Matt and Hall, Simon and Libberton, Paula and Purvis, Francesca},
	title = {Could artificial intelligence write mental health nursing care plans?},
	year = {2024},
	journal = {Journal of Psychiatric and Mental Health Nursing},
	volume = {31},
	number = {1},
	pages = {79 – 86},
	doi = {10.1111/jpm.12965},
	url = {https://www.scopus.com/inward/record.uri?eid=2-s2.0-85167356805&doi=10.1111%2fjpm.12965&partnerID=40&md5=fa6e3ad190e4fc358c85bf39cd2c1492},
	type = {Article},
	publication_stage = {Final},
	source = {Scopus},
	note = {Cited by: 13; All Open Access, Hybrid Gold Open Access}
}

@ARTICLE{Weeks2023,
	author = {Weeks, Rose and Sangha, Pooja and Cooper, Lyra and Sedoc, João and White, Sydney and Gretz, Shai and Toledo, Assaf and Lahav, Dan and Hartner, Anna-Maria and Martin, Nina M. and Lee, Jae Hyoung and Slonim, Noam and Bar-Zeev, Naor},
	title = {Usability and Credibility of a COVID-19 Vaccine Chatbot for Young Adults and Health Workers in the United States: Formative Mixed Methods Study},
	year = {2023},
	journal = {JMIR Human Factors},
	volume = {10},
	doi = {10.2196/40533},
	url = {https://www.scopus.com/inward/record.uri?eid=2-s2.0-85149883396&doi=10.2196%2f40533&partnerID=40&md5=b55f5152bdb85ee3dd34f1440ec5bd36},
	type = {Article},
	publication_stage = {Final},
	source = {Scopus},
	note = {Cited by: 13; All Open Access, Gold Open Access, Green Open Access}
}

@ARTICLE{Dharrao20241,
	author = {Dharrao, Deepak and Gite, Shilpa},
	title = {TherapyBot: a chatbot for mental well-being using transformers},
	year = {2024},
	journal = {International Journal of Advances in Applied Sciences},
	volume = {13},
	number = {1},
	pages = {1 – 12},
	doi = {10.11591/ijaas.v13.i1.pp1-12},
	url = {https://www.scopus.com/inward/record.uri?eid=2-s2.0-85192452835&doi=10.11591%2fijaas.v13.i1.pp1-12&partnerID=40&md5=31db1a45fa2e0557ebf9213b19f27761},
	type = {Article},
	publication_stage = {Final},
	source = {Scopus},
	note = {Cited by: 0; All Open Access, Gold Open Access}
}

@CONFERENCE{Crasto2021,
	author = {Crasto, Reuben and Dias, Lance and Miranda, Dominic and Kayande, Deepali},
	title = {CareBot: A mental health chatbot},
	year = {2021},
	journal = {2021 2nd International Conference for Emerging Technology, INCET 2021},
	doi = {10.1109/INCET51464.2021.9456326},
	url = {https://www.scopus.com/inward/record.uri?eid=2-s2.0-85113330019&doi=10.1109%2fINCET51464.2021.9456326&partnerID=40&md5=e31b28d5eedd570e423d1cad8051cfe2},
	type = {Conference paper},
	publication_stage = {Final},
	source = {Scopus},
	note = {Cited by: 12}
}

@CONFERENCE{Beredo2022300,
	author = {Beredo, Jackylyn L. and Ong, Ethel C.},
	title = {A Hybrid Response Generation Model for an Empathetic Conversational Agent},
	year = {2022},
	journal = {2022 International Conference on Asian Language Processing, IALP 2022},
	pages = {300 – 305},
	doi = {10.1109/IALP57159.2022.9961311},
	url = {https://www.scopus.com/inward/record.uri?eid=2-s2.0-85143975338&doi=10.1109%2fIALP57159.2022.9961311&partnerID=40&md5=8c08772293433af0fdda8397398031c6},
	type = {Conference paper},
	publication_stage = {Final},
	source = {Scopus},
	note = {Cited by: 2}
}

@CONFERENCE{Saha2021,
	author = {Saha, Tulika and Chopra, Saraansh and Saha, Sriparna and Bhattacharyya, Pushpak and Kumar, Pankaj},
	title = {A Large-Scale Dataset for Motivational Dialogue System: An Application of Natural Language Generation to Mental Health},
	year = {2021},
	journal = {Proceedings of the International Joint Conference on Neural Networks},
	volume = {2021-July},
	doi = {10.1109/IJCNN52387.2021.9533924},
	url = {https://www.scopus.com/inward/record.uri?eid=2-s2.0-85116425742&doi=10.1109%2fIJCNN52387.2021.9533924&partnerID=40&md5=3802fc6732ad00fffa45fb152a446f73},
	type = {Conference paper},
	publication_stage = {Final},
	source = {Scopus},
	note = {Cited by: 13}
}

@CONFERENCE{Lubis2018876,
	author = {Lubis, Nurul and Sakti, Sakriani and Yoshino, Koichiro and Nakamura, Satoshi},
	title = {Optimizing Neural Response Generator with Emotional Impact Information},
	year = {2018},
	journal = {2018 IEEE Spoken Language Technology Workshop, SLT 2018 - Proceedings},
	pages = {876 – 883},
	doi = {10.1109/SLT.2018.8639613},
	url = {https://www.scopus.com/inward/record.uri?eid=2-s2.0-85063106395&doi=10.1109%2fSLT.2018.8639613&partnerID=40&md5=8ec60dfd34c18b115a47f74efa83e90d},
	type = {Conference paper},
	publication_stage = {Final},
	source = {Scopus},
	note = {Cited by: 1}
}

@ARTICLE{Brännström2024,
	author = {Brännström, Andreas and Wester, Joel and Nieves, Juan Carlos},
	title = {A formal understanding of computational empathy in interactive agents},
	year = {2024},
	journal = {Cognitive Systems Research},
	volume = {85},
	doi = {10.1016/j.cogsys.2023.101203},
	url = {https://www.scopus.com/inward/record.uri?eid=2-s2.0-85184060695&doi=10.1016%2fj.cogsys.2023.101203&partnerID=40&md5=f39ef77b417fb8306c8e36f716cd5819},
	type = {Article},
	publication_stage = {Final},
	source = {Scopus},
	note = {Cited by: 0; All Open Access, Hybrid Gold Open Access}
}

@ARTICLE{Farrand2024,
	author = {Farrand, Paul and Raue, Patrick J. and Ward, Earlise and Repper, Dean and Areán, Patricia},
	title = {Use and Engagement With Low-Intensity Cognitive Behavioral Therapy Techniques Used Within an App to Support Worry Management: Content Analysis of Log Data},
	year = {2024},
	journal = {JMIR mHealth and uHealth},
	volume = {12},
	number = {1},
	doi = {10.2196/47321},
	url = {https://www.scopus.com/inward/record.uri?eid=2-s2.0-85182091249&doi=10.2196%2f47321&partnerID=40&md5=40176e9c4257eb8b48b294f57ee4233c},
	type = {Article},
	publication_stage = {Final},
	source = {Scopus},
	note = {Cited by: 0; All Open Access, Green Open Access}
}

@ARTICLE{Williams2021,
	author = {Williams, Ruth and Hopkins, Sarah and Frampton, Chris and Holt-Quick, Chester and Merry, Sally Nicola and Stasiak, Karolina},
	title = {21-day stress detox: Open trial of a universal well-being chatbot for young adults},
	year = {2021},
	journal = {Social Sciences},
	volume = {10},
	number = {11},
	doi = {10.3390/socsci10110416},
	url = {https://www.scopus.com/inward/record.uri?eid=2-s2.0-85121629602&doi=10.3390%2fsocsci10110416&partnerID=40&md5=1bc3103d9d3d59d93f9c8270c999933e},
	type = {Article},
	publication_stage = {Final},
	source = {Scopus},
	note = {Cited by: 11; All Open Access, Gold Open Access}
}

@ARTICLE{Garcia-Velez2024389,
	author = {Garcia-Velez, Roberto and Quisi-Peralta, Diego and Salgado-Guerrero, Juan and Cárdenas-Arichábala, Tracy and Paguay-Palaguachi, Luis and Angamarca-Deleg, Verónica},
	title = {Test Case of a Chatbot-Based Virtual Assistant to Identify and Classify Cases of Harassment Within the University and Refer to the Student Welfare Department Using GPT-3},
	year = {2024},
	journal = {Lecture Notes in Networks and Systems},
	volume = {932 LNNS},
	pages = {389 – 397},
	doi = {10.1007/978-3-031-54235-0_35},
	url = {https://www.scopus.com/inward/record.uri?eid=2-s2.0-85187791340&doi=10.1007%2f978-3-031-54235-0_35&partnerID=40&md5=a4a060cc595ec5afe874aa90c8a629f7},
	type = {Conference paper},
	publication_stage = {Final},
	source = {Scopus},
	note = {Cited by: 0}
}

@ARTICLE{Gabrielli2020,
	author = {Gabrielli, Silvia and Rizzi, Silvia and Carbone, Sara and Donisi, Valeria},
	title = {A chatbot-based coaching intervention for adolescents to promote life skills: Pilot study},
	year = {2020},
	journal = {JMIR Human Factors},
	volume = {7},
	number = {1},
	doi = {10.2196/16762},
	url = {https://www.scopus.com/inward/record.uri?eid=2-s2.0-85091105700&doi=10.2196%2f16762&partnerID=40&md5=ae2d212df895acbb8a55ea91ab352a2a},
	type = {Article},
	publication_stage = {Final},
	source = {Scopus},
	note = {Cited by: 45; All Open Access, Gold Open Access, Green Open Access}
}

@ARTICLE{Ahmad2022923,
	author = {Ahmad, Rangina and Siemon, Dominik and Gnewuch, Ulrich and Robra-Bissantz, Susanne},
	title = {Designing Personality-Adaptive Conversational Agents for Mental Health Care},
	year = {2022},
	journal = {Information Systems Frontiers},
	volume = {24},
	number = {3},
	pages = {923 – 943},
	doi = {10.1007/s10796-022-10254-9},
	url = {https://www.scopus.com/inward/record.uri?eid=2-s2.0-85126123152&doi=10.1007%2fs10796-022-10254-9&partnerID=40&md5=b9a5e6546d130be01a9e0b0b184d2700},
	type = {Article},
	publication_stage = {Final},
	source = {Scopus},
	note = {Cited by: 42; All Open Access, Hybrid Gold Open Access}
}

@ARTICLE{Ly201739,
	author = {Ly, Kien Hoa and Ly, Ann-Marie and Andersson, Gerhard},
	title = {A fully automated conversational agent for promoting mental well-being: A pilot RCT using mixed methods},
	year = {2017},
	journal = {Internet Interventions},
	volume = {10},
	pages = {39 – 46},
	doi = {10.1016/j.invent.2017.10.002},
	url = {https://www.scopus.com/inward/record.uri?eid=2-s2.0-85032031216&doi=10.1016%2fj.invent.2017.10.002&partnerID=40&md5=63c74351b255bbd427a9f15ce03b4b47},
	type = {Article},
	publication_stage = {Final},
	source = {Scopus},
	note = {Cited by: 228; All Open Access, Gold Open Access}
}

@ARTICLE{Cheng2024140,
	author = {Cheng, Sunny Chieh and Kopelovich, Sarah and Si, Dong and Divina, Myra and Gao, Ningjun Serene and Wang, Mia Yunqi and Kim, Jamie Jaesook and Li, Ziyi and Blank, Jennifer and Brian, Rachel and Turkington, Douglas},
	title = {Co-Production of a Cognitive Behavioral Therapy Digital Platform for Families of Individuals Impacted by Psychosis},
	year = {2024},
	journal = {Journal of Technology in Behavioral Science},
	volume = {9},
	number = {1},
	pages = {140 – 148},
	doi = {10.1007/s41347-023-00378-3},
	url = {https://www.scopus.com/inward/record.uri?eid=2-s2.0-85181676797&doi=10.1007%2fs41347-023-00378-3&partnerID=40&md5=d5e496007b393e3c9b0fa54583f702f2},
	type = {Article},
	publication_stage = {Final},
	source = {Scopus},
	note = {Cited by: 0}
}

@ARTICLE{Karkosz2024,
	author = {Karkosz, Stanisław and Szymański, Robert and Sanna, Katarzyna and Michałowski, Jarosław},
	title = {Effectiveness of a Web-based and Mobile Therapy Chatbot on Anxiety and Depressive Symptoms in Subclinical Young Adults: Randomized Controlled Trial},
	year = {2024},
	journal = {JMIR Formative Research},
	volume = {8},
	doi = {10.2196/47960},
	url = {https://www.scopus.com/inward/record.uri?eid=2-s2.0-85191226596&doi=10.2196%2f47960&partnerID=40&md5=dbb7c054cc6ea0fc204fba0cb5a244e2},
	type = {Article},
	publication_stage = {Final},
	source = {Scopus},
	note = {Cited by: 2; All Open Access, Gold Open Access}
}

@ARTICLE{Elyoseph2023,
	author = {Elyoseph, Zohar and Hadar-Shoval, Dorit and Asraf, Kfir and Lvovsky, Maya},
	title = {ChatGPT outperforms humans in emotional awareness evaluations},
	year = {2023},
	journal = {Frontiers in Psychology},
	volume = {14},
	doi = {10.3389/fpsyg.2023.1199058},
	url = {https://www.scopus.com/inward/record.uri?eid=2-s2.0-85161937174&doi=10.3389%2ffpsyg.2023.1199058&partnerID=40&md5=d9e02bf0a352a48dc6b683bf9e0f2186},
	type = {Article},
	publication_stage = {Final},
	source = {Scopus},
	note = {Cited by: 60; All Open Access, Gold Open Access}
}

@CONFERENCE{Mansoori2022,
	author = {Mansoori, Madiha and Maliwal, Hrishil and Kotian, Sharvil and Kenkre, Hersh and Saha, Ishani and Mishra, Payal},
	title = {A Systematic Survey on Computational agents for Mental Health Aid},
	year = {2022},
	journal = {2022 IEEE 7th International conference for Convergence in Technology, I2CT 2022},
	doi = {10.1109/I2CT54291.2022.9824269},
	url = {https://www.scopus.com/inward/record.uri?eid=2-s2.0-85135620071&doi=10.1109%2fI2CT54291.2022.9824269&partnerID=40&md5=9635de90a915b873bb742b695770031e},
	type = {Conference paper},
	publication_stage = {Final},
	source = {Scopus},
	note = {Cited by: 3}
}

@ARTICLE{Zhang2011412,
	author = {Zhang, Li},
	title = {Exploration on context-sensitive affect sensing in an intelligent agent},
	year = {2011},
	journal = {Lecture Notes in Computer Science (including subseries Lecture Notes in Artificial Intelligence and Lecture Notes in Bioinformatics)},
	volume = {6895 LNAI},
	pages = {412 – 418},
	doi = {10.1007/978-3-642-23974-8_44},
	url = {https://www.scopus.com/inward/record.uri?eid=2-s2.0-80053184640&doi=10.1007%2f978-3-642-23974-8_44&partnerID=40&md5=03ef8100df37d2ca252c86b5825b09e8},
	type = {Conference paper},
	publication_stage = {Final},
	source = {Scopus},
	note = {Cited by: 0}
}

@CONFERENCE{Hayashi2024521,
	author = {Hayashi, Yugo and Kiuchi, Keita},
	title = {Towards Developing an Active Listening Counseling Robot for Multiple Generations A Text Mining Study on Emotional Expression of the Elderly and the Young},
	year = {2024},
	journal = {ACM/IEEE International Conference on Human-Robot Interaction},
	pages = {521 – 525},
	doi = {10.1145/3610978.3640733},
	url = {https://www.scopus.com/inward/record.uri?eid=2-s2.0-85188076780&doi=10.1145%2f3610978.3640733&partnerID=40&md5=fc9a7b9a85ea35db49fec94f64ef51ac},
	type = {Conference paper},
	publication_stage = {Final},
	source = {Scopus},
	note = {Cited by: 0}
}

@ARTICLE{Hopman2024161,
	author = {Hopman, Katherine and Richards, Deborah and Norberg, Melissa N.},
	title = {An Embodied Conversational Agent to Support Wellbeing After Injury: Insights from a Stakeholder Inclusive Design Approach},
	year = {2024},
	journal = {Lecture Notes in Computer Science (including subseries Lecture Notes in Artificial Intelligence and Lecture Notes in Bioinformatics)},
	volume = {14636 LNCS},
	pages = {161 – 175},
	doi = {10.1007/978-3-031-58226-4_13},
	url = {https://www.scopus.com/inward/record.uri?eid=2-s2.0-85192174788&doi=10.1007%2f978-3-031-58226-4_13&partnerID=40&md5=ad12827086780462edc4f45037da3795},
	type = {Conference paper},
	publication_stage = {Final},
	source = {Scopus},
	note = {Cited by: 0}
}

@ARTICLE{Torkamaan2023,
	author = {Torkamaan, Helma},
	title = {Mood Measurement on Smartphones: Which Measure, Which Design?},
	year = {2023},
	journal = {Proceedings of the ACM on Interactive, Mobile, Wearable and Ubiquitous Technologies},
	volume = {7},
	number = {1},
	doi = {10.1145/3580864},
	url = {https://www.scopus.com/inward/record.uri?eid=2-s2.0-85152489968&doi=10.1145%2f3580864&partnerID=40&md5=858bd091401500665e0422264ba9fe13},
	type = {Article},
	publication_stage = {Final},
	source = {Scopus},
	note = {Cited by: 1; All Open Access, Bronze Open Access}
}

@ARTICLE{Jabir2024,
	author = {Jabir, Ahmad Ishqi and Lin, Xiaowen and Martinengo, Laura and Sharp, Gemma and Theng, Yin-Leng and Car, Lorainne Tudor},
	title = {Attrition in Conversational Agent-Delivered Mental Health Interventions: Systematic Review and Meta-Analysis},
	year = {2024},
	journal = {Journal of Medical Internet Research},
	volume = {26},
	number = {1},
	doi = {10.2196/48168},
	url = {https://www.scopus.com/inward/record.uri?eid=2-s2.0-85186391608&doi=10.2196%2f48168&partnerID=40&md5=9c5266d8c81c5bff631e2e3e588965af},
	type = {Review},
	publication_stage = {Final},
	source = {Scopus},
	note = {Cited by: 2; All Open Access, Gold Open Access}
}

@CONFERENCE{Liao2023,
	author = {Liao, Ting and Yan, Bei},
	title = {LET'S CHAT IF YOU ARE UNHAPPY - THE EFFECT OF EMOTIONS ON INTERACTION EXPERIENCE AND TRUST TOWARD EMPATHETIC CHATBOTS},
	year = {2023},
	journal = {Proceedings of the ASME Design Engineering Technical Conference},
	volume = {3B},
	doi = {10.1115/DETC2023-115318},
	url = {https://www.scopus.com/inward/record.uri?eid=2-s2.0-85179134798&doi=10.1115%2fDETC2023-115318&partnerID=40&md5=35b9671dc97d880bdf74c1fe70ea1587},
	type = {Conference paper},
	publication_stage = {Final},
	source = {Scopus},
	note = {Cited by: 0}
}

@CONFERENCE{Maharjan2021,
	author = {Maharjan, Raju and Rohani, Darius Adam and Bækgaard, Per and Bardram, Jakob and Doherty, Kevin},
	title = {Can we talk? Design Implications for the Questionnaire-Driven Self-Report of Health and Wellbeing via Conversational Agent},
	year = {2021},
	journal = {ACM International Conference Proceeding Series},
	doi = {10.1145/3469595.3469600},
	url = {https://www.scopus.com/inward/record.uri?eid=2-s2.0-85112247533&doi=10.1145%2f3469595.3469600&partnerID=40&md5=beddb18cd82a7ec9dcd43a1f6c548875},
	type = {Conference paper},
	publication_stage = {Final},
	source = {Scopus},
	note = {Cited by: 13; All Open Access, Bronze Open Access, Green Open Access}
}

@CONFERENCE{Bravo2020941,
	author = {Bravo, Sean Latrelle and Herrera, Cedric Jose and Valdez, Edward Carlo and Poliquit, Klint John and Ureta, Jennifer and Cu, Jocelynn and Azcarraga, Judith and Rivera, Joanna Pauline},
	title = {CATe: An embodied conversational agent for the elderly},
	year = {2020},
	journal = {ICAART 2020 - Proceedings of the 12th International Conference on Agents and Artificial Intelligence},
	volume = {2},
	pages = {941 – 948},
	url = {https://www.scopus.com/inward/record.uri?eid=2-s2.0-85083088113&partnerID=40&md5=871290b281dd39b128012e101ae5dd3a},
	type = {Conference paper},
	publication_stage = {Final},
	source = {Scopus},
	note = {Cited by: 8}
}

@ARTICLE{Lisetti2013,
	author = {Lisetti, Christine and Amini, Reza and Yasavur, Ugan and Rishe, Naphtali},
	title = {I can help you change! An empathic virtual agent delivers behavior change health interventions},
	year = {2013},
	journal = {ACM Transactions on Management Information Systems},
	volume = {4},
	number = {4},
	doi = {10.1145/2544103},
	url = {https://www.scopus.com/inward/record.uri?eid=2-s2.0-84891769876&doi=10.1145%2f2544103&partnerID=40&md5=01cfe4a0f4c2d74096db54bbf9cdad9d},
	type = {Article},
	publication_stage = {Final},
	source = {Scopus},
	note = {Cited by: 174}
}

@CONFERENCE{Schwartz2022139,
	author = {Schwartz, R.X. and Ramanan, Aparna and Patel, Disha and Lynch, Annabel and Baee, Sonia and Barnes, Laura},
	title = {DARA: Development of a Chatbot Support System for an Anxiety Reduction Digital Intervention},
	year = {2022},
	journal = {2022 Systems and Information Engineering Design Symposium, SIEDS 2022},
	pages = {139 – 144},
	doi = {10.1109/SIEDS55548.2022.9799419},
	url = {https://www.scopus.com/inward/record.uri?eid=2-s2.0-85134344041&doi=10.1109%2fSIEDS55548.2022.9799419&partnerID=40&md5=aaacd93b5eaceb0e4c0014e51345f276},
	type = {Conference paper},
	publication_stage = {Final},
	source = {Scopus},
	note = {Cited by: 2}
}

@CONFERENCE{Valtolina2021,
	author = {Valtolina, Stefano and Hu, Liliana},
	title = {Charlie: A chatbot to improve the elderly quality of life and to make them more active to fight their sense of loneliness},
	year = {2021},
	journal = {ACM International Conference Proceeding Series},
	doi = {10.1145/3464385.3464726},
	url = {https://www.scopus.com/inward/record.uri?eid=2-s2.0-85112685594&doi=10.1145%2f3464385.3464726&partnerID=40&md5=ba8f7b642786a38d483a110fa4baf129},
	type = {Conference paper},
	publication_stage = {Final},
	source = {Scopus},
	note = {Cited by: 13}
}

@ARTICLE{Cameron2019121,
	author = {Cameron, Gillian and Cameron, David and Megaw, Gavin and Bond, Raymond and Mulvenna, Maurice and O’Neill, Siobhan and Armour, Cherie and McTear, Michael},
	title = {Assessing the usability of a chatbot for mental health care},
	year = {2019},
	journal = {Lecture Notes in Computer Science (including subseries Lecture Notes in Artificial Intelligence and Lecture Notes in Bioinformatics)},
	volume = {11551 LNCS},
	pages = {121 – 132},
	doi = {10.1007/978-3-030-17705-8_11},
	url = {https://www.scopus.com/inward/record.uri?eid=2-s2.0-85065299319&doi=10.1007%2f978-3-030-17705-8_11&partnerID=40&md5=4a80150d390ef49ba701218589d1e2cb},
	type = {Conference paper},
	publication_stage = {Final},
	source = {Scopus},
	note = {Cited by: 53}
}

@ARTICLE{Escobar-Viera2023,
	author = {Escobar-Viera, César G. and Porta, Giovanna and Coulter, Robert W.S. and Martina, Jamie and Goldbach, Jeremy and Rollman, Bruce L.},
	title = {A chatbot-delivered intervention for optimizing social media use and reducing perceived isolation among rural-living LGBTQ+ youth: Development, acceptability, usability, satisfaction, and utility},
	year = {2023},
	journal = {Internet Interventions},
	volume = {34},
	doi = {10.1016/j.invent.2023.100668},
	url = {https://www.scopus.com/inward/record.uri?eid=2-s2.0-85170643995&doi=10.1016%2fj.invent.2023.100668&partnerID=40&md5=7b77be8ea3375f18d2d90b0c4762534f},
	type = {Article},
	publication_stage = {Final},
	source = {Scopus},
	note = {Cited by: 5; All Open Access, Gold Open Access}
}

@CONFERENCE{Boyd2022,
	author = {Boyd, Kyle and Potts, Courtney and Bond, Raymond and Mulvenna, Maurice and Broderick, Thomas and Burns, Con and Bickerdike, Andrea and Mctear, Mike and Kostenius, Catrine and Vakaloudis, Alex and Dhanapala, Indika and Ennis, Edel and Booth, Fred},
	title = {Usability testing and trust analysis of a mental health and wellbeing chatbot},
	year = {2022},
	journal = {ACM International Conference Proceeding Series},
	doi = {10.1145/3552327.3552348},
	url = {https://www.scopus.com/inward/record.uri?eid=2-s2.0-85139957448&doi=10.1145%2f3552327.3552348&partnerID=40&md5=0ac20b4bf1e2d022db0695ec8cb6b254},
	type = {Conference paper},
	publication_stage = {Final},
	source = {Scopus},
	note = {Cited by: 1; All Open Access, Bronze Open Access, Green Open Access}
}

@ARTICLE{Ellis-Brush2021187,
	author = {Ellis-Brush, Kevin},
	title = {Augmenting Coaching Practice through Digital Methods},
	year = {2021},
	journal = {International Journal of Evidence Based Coaching and Mentoring},
	pages = {187 – 197},
	doi = {10.24384/er2p-4857},
	url = {https://www.scopus.com/inward/record.uri?eid=2-s2.0-85108668332&doi=10.24384%2fer2p-4857&partnerID=40&md5=6f38a055245edc12da802597aa83e6dd},
	type = {Article},
	publication_stage = {Final},
	source = {Scopus},
	note = {Cited by: 8}
}

@ARTICLE{Drouin2022,
	author = {Drouin, Michelle and Sprecher, Susan and Nicola, Robert and Perkins, Taylor},
	title = {Is chatting with a sophisticated chatbot as good as chatting online or FTF with a stranger?},
	year = {2022},
	journal = {Computers in Human Behavior},
	volume = {128},
	doi = {10.1016/j.chb.2021.107100},
	url = {https://www.scopus.com/inward/record.uri?eid=2-s2.0-85120172387&doi=10.1016%2fj.chb.2021.107100&partnerID=40&md5=bc990236c0350535d7b9a0794510d472},
	type = {Article},
	publication_stage = {Final},
	source = {Scopus},
	note = {Cited by: 23}
}

@ARTICLE{Rodríguez-Martínez2024,
	author = {Rodríguez-Martínez, Antonia and Amezcua-Aguilar, Teresa and Cortés-Moreno, Javier and Jiménez-Delgado, Juan José},
	title = {Qualitative Analysis of Conversational Chatbots to Alleviate Loneliness in Older Adults as a Strategy for Emotional Health},
	year = {2024},
	journal = {Healthcare (Switzerland)},
	volume = {12},
	number = {1},
	doi = {10.3390/healthcare12010062},
	url = {https://www.scopus.com/inward/record.uri?eid=2-s2.0-85181959205&doi=10.3390%2fhealthcare12010062&partnerID=40&md5=99c4b33891d4867be10c56ab8e41e4cd},
	type = {Article},
	publication_stage = {Final},
	source = {Scopus},
	note = {Cited by: 0; All Open Access, Gold Open Access}
}

@CONFERENCE{Niess2018916,
	author = {Niess, Jasmin and Diefenbach, Sarah and Platz, Axel},
	title = {Moving beyond assistance: Psychological qualities of digital companions},
	year = {2018},
	journal = {ACM International Conference Proceeding Series},
	pages = {916 – 921},
	doi = {10.1145/3240167.3240240},
	url = {https://www.scopus.com/inward/record.uri?eid=2-s2.0-85056580141&doi=10.1145%2f3240167.3240240&partnerID=40&md5=50bba924b5537f6fc62ec53c80f8de84},
	type = {Conference paper},
	publication_stage = {Final},
	source = {Scopus},
	note = {Cited by: 6}
}

@ARTICLE{Andrade-Arenas2024114,
	author = {Andrade-Arenas, Laberiano and Yactayo-Arias, Cesar and Pucuhuayla-Revatta, Félix},
	title = {Therapy and Emotional Support through a Chatbot},
	year = {2024},
	journal = {International journal of online and biomedical engineering},
	volume = {20},
	number = {2},
	pages = {114 – 130},
	doi = {10.3991/ijoe.v20i02.45377},
	url = {https://www.scopus.com/inward/record.uri?eid=2-s2.0-85186860956&doi=10.3991%2fijoe.v20i02.45377&partnerID=40&md5=9a6e05588a86eac015e8e61e30d296a3},
	type = {Article},
	publication_stage = {Final},
	source = {Scopus},
	note = {Cited by: 0; All Open Access, Gold Open Access}
}

@CONFERENCE{Ahmad2021,
	author = {Ahmad, Rangina and Siemon, Dominik and Gnewuch, Ulrich and Robra-Bissantz, Susanne},
	title = {The benefits and caveats of personality-adaptive conversational agents in mental health care},
	year = {2021},
	journal = {27th Annual Americas Conference on Information Systems, AMCIS 2021},
	url = {https://www.scopus.com/inward/record.uri?eid=2-s2.0-85118663803&partnerID=40&md5=7ffb768870bc878913dd7bfeba7c5725},
	type = {Conference paper},
	publication_stage = {Final},
	source = {Scopus},
	note = {Cited by: 7}
}

@ARTICLE{Benke2020,
	author = {Benke, Ivo and Knierim, Michael Thomas and Maedche, Alexander},
	title = {Chatbot-based Emotion Management for Distributed Teams: A Participatory Design Study},
	year = {2020},
	journal = {Proceedings of the ACM on Human-Computer Interaction},
	volume = {4},
	number = {CSCW2},
	doi = {10.1145/3415189},
	url = {https://www.scopus.com/inward/record.uri?eid=2-s2.0-85094217965&doi=10.1145%2f3415189&partnerID=40&md5=1e87147af4add6ffa6e7264a21303241},
	type = {Article},
	publication_stage = {Final},
	source = {Scopus},
	note = {Cited by: 38}
}

@CONFERENCE{Fadhil2018,
	author = {Fadhil, Ahmed and Schiavo, Gianluca and Wang, Yunlong and Yilma, Bereket A.},
	title = {The effect of emojis when interacting with conversationinterface assisted health coaching system},
	year = {2018},
	journal = {PervasiveHealth: Pervasive Computing Technologies for Healthcare},
	doi = {10.1145/3240925.3240965},
	url = {https://www.scopus.com/inward/record.uri?eid=2-s2.0-85116372539&doi=10.1145%2f3240925.3240965&partnerID=40&md5=53cbd3b18d253661f21e444b94a76e9d},
	type = {Conference paper},
	publication_stage = {Final},
	source = {Scopus},
	note = {Cited by: 17; All Open Access, Green Open Access}
}

@ARTICLE{Li2024722,
	author = {Li, Lin and Peng, Wei and Rheu, Minjin M.J.},
	title = {Factors Predicting Intentions of Adoption and Continued Use of Artificial Intelligence Chatbots for Mental Health: Examining the Role of UTAUT Model, Stigma, Privacy Concerns, and Artificial Intelligence Hesitancy},
	year = {2024},
	journal = {Telemedicine and e-Health},
	volume = {30},
	number = {3},
	pages = {722 – 730},
	doi = {10.1089/tmj.2023.0313},
	url = {https://www.scopus.com/inward/record.uri?eid=2-s2.0-85174635368&doi=10.1089%2ftmj.2023.0313&partnerID=40&md5=0a1a73e07bd4760ca0a840cdf7b9ac4c},
	type = {Article},
	publication_stage = {Final},
	source = {Scopus},
	note = {Cited by: 7}
}

@ARTICLE{Bendig2021,
	author = {Bendig, Eileen and Erb, Benjamin and Meißner, Dominik and Bauereiß, Natalie and Baumeister, Harald},
	title = {Feasibility of a Software agent providing a brief Intervention for Self-help to Uplift psychological wellbeing (“SISU”). A single-group pretest-posttest trial investigating the potential of SISU to act as therapeutic agent},
	year = {2021},
	journal = {Internet Interventions},
	volume = {24},
	doi = {10.1016/j.invent.2021.100377},
	url = {https://www.scopus.com/inward/record.uri?eid=2-s2.0-85102639630&doi=10.1016%2fj.invent.2021.100377&partnerID=40&md5=13531860d1b6fd8d3c6e1139ef92169a},
	type = {Article},
	publication_stage = {Final},
	source = {Scopus},
	note = {Cited by: 16; All Open Access, Gold Open Access, Green Open Access}
}

@CONFERENCE{Horii2019190,
	author = {Horii, Tsubasa and Sakurai, Yoshitaka and Sakurai, Eriko and Tsuruta, Setsuo and Knauf, Rainer and Damiani, Ernesto and Kutics, Andrea},
	title = {More general evaluation of a client-centered counseling agent},
	year = {2019},
	journal = {Proceedings - 2019 IEEE World Congress on Services, SERVICES 2019},
	pages = {190 – 196},
	doi = {10.1109/SERVICES.2019.00052},
	url = {https://www.scopus.com/inward/record.uri?eid=2-s2.0-85072766284&doi=10.1109%2fSERVICES.2019.00052&partnerID=40&md5=de68134c0438e30744151e0dc5f2f03a},
	type = {Conference paper},
	publication_stage = {Final},
	source = {Scopus},
	note = {Cited by: 2}
}

@ARTICLE{Kaywan2023,
	author = {Kaywan, Payam and Ahmed, Khandakar and Ibaida, Ayman and Miao, Yuan and Gu, Bruce},
	title = {Early detection of depression using a conversational AI bot: A non-clinical trial},
	year = {2023},
	journal = {PLoS ONE},
	volume = {18},
	number = {2 February},
	doi = {10.1371/journal.pone.0279743},
	url = {https://www.scopus.com/inward/record.uri?eid=2-s2.0-85147457613&doi=10.1371%2fjournal.pone.0279743&partnerID=40&md5=d9f40464198f91c11572b1ebd17fdeab},
	type = {Article},
	publication_stage = {Final},
	source = {Scopus},
	note = {Cited by: 17; All Open Access, Gold Open Access, Green Open Access}
}

@CONFERENCE{Hakani2022796,
	author = {Hakani, Riddhi and Patil, Samiksha and Patil, Sakshi and Jhunjhunwala, Siddhi and Deulkar, Khushali},
	title = {Revivify: A Depression Detection and Control System using Tweets and Automated Chatbot},
	year = {2022},
	journal = {Proceedings - 2022 IEEE World Conference on Applied Intelligence and Computing, AIC 2022},
	pages = {796 – 801},
	doi = {10.1109/AIC55036.2022.9848978},
	url = {https://www.scopus.com/inward/record.uri?eid=2-s2.0-85137842722&doi=10.1109%2fAIC55036.2022.9848978&partnerID=40&md5=8ff520d048ca8d797f0fc048d358f0f9},
	type = {Conference paper},
	publication_stage = {Final},
	source = {Scopus},
	note = {Cited by: 0}
}

@ARTICLE{Kaywan2021193,
	author = {Kaywan, Payam and Ahmed, Khandakar and Miao, Yuan and Ibaida, Ayman and Gu, Bruce},
	title = {DEPRA: An Early Depression Detection Analysis Chatbot},
	year = {2021},
	journal = {Lecture Notes in Computer Science (including subseries Lecture Notes in Artificial Intelligence and Lecture Notes in Bioinformatics)},
	volume = {13079 LNCS},
	pages = {193 – 204},
	doi = {10.1007/978-3-030-90885-0_18},
	url = {https://www.scopus.com/inward/record.uri?eid=2-s2.0-85120054501&doi=10.1007%2f978-3-030-90885-0_18&partnerID=40&md5=5021898037dcbc087363660fd52b0b02},
	type = {Conference paper},
	publication_stage = {Final},
	source = {Scopus},
	note = {Cited by: 3}
}

@ARTICLE{Blease2024,
	author = {Blease, Charlotte and Worthen, Abigail and Torous, John},
	title = {Psychiatrists’ experiences and opinions of generative artificial intelligence in mental healthcare: An online mixed methods survey},
	year = {2024},
	journal = {Psychiatry Research},
	volume = {333},
	doi = {10.1016/j.psychres.2024.115724},
	url = {https://www.scopus.com/inward/record.uri?eid=2-s2.0-85184764032&doi=10.1016%2fj.psychres.2024.115724&partnerID=40&md5=b24349aa59eaecc1af127264d01a91b5},
	type = {Article},
	publication_stage = {Final},
	source = {Scopus},
	note = {Cited by: 5; All Open Access, Hybrid Gold Open Access}
}

@Article{Wrightson-Hester2023,
author="Wrightson-Hester, Aimee-Rose
and Anderson, Georgia
and Dunstan, Joel
and McEvoy, Peter M
and Sutton, Christopher J
and Myers, Bronwyn
and Egan, Sarah
and Tai, Sara
and Johnston-Hollitt, Melanie
and Chen, Wai
and Gedeon, Tom
and Mansell, Warren",
title="An Artificial Therapist (Manage Your Life Online) to Support the Mental Health of Youth: Co-Design and Case Series",
journal="JMIR Hum Factors",
year="2023",
month="Jul",
day="21",
volume="10",
pages="e46849",
issn="2292-9495",
doi="10.2196/46849",
url="https://humanfactors.jmir.org/2023/1/e46849",
url="https://doi.org/10.2196/46849",
url="http://www.ncbi.nlm.nih.gov/pubmed/37477969"
}

@ARTICLE{Nicol2022,
	author = {Nicol, Ginger and Wang, Ruoyun and Graham, Sharon and Dodd, Sherry and Garbutt, Jane},
	title = {Chatbot-Delivered Cognitive Behavioral Therapy in Adolescents With Depression and Anxiety During the COVID-19 Pandemic: Feasibility and Acceptability Study},
	year = {2022},
	journal = {JMIR Formative Research},
	volume = {6},
	number = {11},
	doi = {10.2196/40242},
	url = {https://www.scopus.com/inward/record.uri?eid=2-s2.0-85144759725&doi=10.2196%2f40242&partnerID=40&md5=ca4638a0c3dd63ccdb06661b233b8c27},
	type = {Article},
	publication_stage = {Final},
	source = {Scopus},
	note = {Cited by: 23; All Open Access, Gold Open Access}
}

@BOOK{Solanki2023143,
	author = {Solanki, Nisha and Singh, Nidhi and Garg, Disha},
	title = {Human mental experience through chatbots: A thematical analysis of human engagement with evidence-based cognitivebehavioral techniques},
	year = {2023},
	journal = {Edge-AI in Healthcare: Trends and Future Perspectives},
	pages = {143 – 158},
	doi = {10.1201/9781003244592-11},
	url = {https://www.scopus.com/inward/record.uri?eid=2-s2.0-85167762978&doi=10.1201%2f9781003244592-11&partnerID=40&md5=df3103f99e55c5c8d8f5f0a0aa7955b0},
	type = {Book chapter},
	publication_stage = {Final},
	source = {Scopus},
	note = {Cited by: 1}
}

@ARTICLE{He2022,
	author = {He, Yuhao and Yang, Li and Zhu, Xiaokun and Wu, Bin and Zhang, Shuo and Qian, Chunlian and Tian, Tian},
	title = {Mental Health Chatbot for Young Adults With Depressive Symptoms During the COVID-19 Pandemic: Single-Blind, Three-Arm Randomized Controlled Trial},
	year = {2022},
	journal = {Journal of Medical Internet Research},
	volume = {24},
	number = {11},
	doi = {10.2196/40719},
	url = {https://www.scopus.com/inward/record.uri?eid=2-s2.0-85142401562&doi=10.2196%2f40719&partnerID=40&md5=f7fae1cbc004a858163af84dd1d79283},
	type = {Article},
	publication_stage = {Final},
	source = {Scopus},
	note = {Cited by: 33; All Open Access, Gold Open Access, Green Open Access}
}

@ARTICLE{Provoost2020,
	author = {Provoost, Simon and Kleiboer, Annet and Ornelas, José and Bosse, Tibor and Ruwaard, Jeroen and Rocha, Artur and Cuijpers, Pim and Riper, Heleen},
	title = {Improving adherence to an online intervention for low mood with a virtual coach: study protocol of a pilot randomized controlled trial},
	year = {2020},
	journal = {Trials},
	volume = {21},
	number = {1},
	doi = {10.1186/s13063-020-04777-2},
	type = {Article},
	publication_stage = {Final},
	source = {Scopus},
	note = {Cited by: 7; All Open Access, Gold Open Access}
}

@ARTICLE{Chou2024,
	author = {Chou, Ya-Hsin and Lin, Chemin and Lee, Shwu-Hua and Lee, Yen-Fen and Cheng, Li-Chen},
	title = {User-Friendly Chatbot to Mitigate the Psychological Stress of Older Adults During the COVID-19 Pandemic: Development and Usability Study},
	year = {2024},
	journal = {JMIR Formative Research},
	volume = {8},
	doi = {10.2196/49462},
	type = {Article},
	publication_stage = {Final},
	source = {Scopus},
	note = {Cited by: 0; All Open Access, Gold Open Access}
}

@ARTICLE{Shah20221229,
	author = {Shah, Jillian and DePietro, Bianca and D'Adamo, Laura and Firebaugh, Marie-Laure and Laing, Olivia and Fowler, Lauren A. and Smolar, Lauren and Sadeh-Sharvit, Shiri and Taylor, C. Barr and Wilfley, Denise E. and Fitzsimmons-Craft, Ellen E.},
	title = {Development and usability testing of a chatbot to promote mental health services use among individuals with eating disorders following screening},
	year = {2022},
	journal = {International Journal of Eating Disorders},
	volume = {55},
	number = {9},
	pages = {1229 – 1244},
	doi = {10.1002/eat.23798},
	type = {Article},
	publication_stage = {Final},
	source = {Scopus},
	note = {Cited by: 21; All Open Access, Green Open Access}
}

@CONFERENCE{Barbosa2022355,
	author = {Barbosa, Marcus and Nakamura, Walter and Valle, Pedro and Guerino, Guilherme C. and Finger, Alice F. and Lunardi, Gabriel M. and Silva, Williamson},
	title = {UX of Chatbots: An Exploratory Study on Acceptance of User Experience Evaluation Methods},
	year = {2022},
	journal = {International Conference on Enterprise Information Systems, ICEIS - Proceedings},
	volume = {2},
	pages = {355 – 363},
	doi = {10.5220/0011090100003179},
	type = {Conference paper},
	publication_stage = {Final},
	source = {Scopus},
	note = {Cited by: 2; All Open Access, Hybrid Gold Open Access}
}

@ARTICLE{Andrews2023418,
	author = {Andrews, Brooke and Klein, Britt and Corboy, Denise and McLaren, Suzanne and Watson, Shaun},
	title = {Video Chat Therapist Assistance in an Adaptive Digital Intervention for Anxiety and Depression: Reflections From Participants and Therapists},
	year = {2023},
	journal = {Professional Psychology: Research and Practice},
	volume = {54},
	number = {6},
	pages = {418 – 429},
	doi = {10.1037/pro0000527},
	type = {Article},
	publication_stage = {Final},
	source = {Scopus},
	note = {Cited by: 0}
}

@ARTICLE{Fitzpatrick2017,
	author = {Fitzpatrick, Kathleen Kara and Darcy, Alison and Vierhile, Molly},
	title = {Delivering cognitive behavior therapy to young adults with symptoms of depression and anxiety using a fully automated conversational agent (Woebot): A randomized controlled trial},
	year = {2017},
	journal = {JMIR Mental Health},
	volume = {4},
	number = {2},
	doi = {10.2196/mental.7785},
	type = {Article},
	publication_stage = {Final},
	source = {Scopus},
	note = {Cited by: 1120; All Open Access, Gold Open Access, Green Open Access}
}

@ARTICLE{Anmella2023,
	author = {Anmella, Gerard and Sanabra, Miriam and Primé-Tous, Mireia and Segú, Xavier and Cavero, Myriam and Morilla, Ivette and Grande, Iria and Ruiz, Victoria and Mas, Ariadna and Martín-Villalba, Inés and Caballo, Alejandro and Julia-Parisad, Esteva and Rodríguez-Rey, Arturo and Piazza, Flavia and José Valdesoiro, Francisco and Rodriguez-Torrella, Claudia and Espinosa, Marta and Virgili, Giulia and Sorroche, Carlota and Ruiz, Alicia and Solanes, Aleix and Radua, Joaquim and Antonieta Also, María and Sant, Elisenda and Murgui, Sandra and Sans-Corrales, Mireia and Young, Allan H. and Vicens, Victor and Blanch, Jordi and Caballeria, Elsa and López-Pelayo, Hugo and López, Clara and Olivé, Victoria and Pujol, Laura and Quesada, Sebastiana and Solé, Brisa and Torrent, Carla and Martínez-Aran, Anabel and Guarch, Joana and Navinés, Ricard and Murru, Andrea and Fico, Giovanna and de Prisco, Michele and Oliva, Vicenzo and Amoretti, Silvia and Pio-Carrino, Casimiro and Fernández-Canseco, María and Villegas, Marta and Vieta, Eduard and Hidalgo-Mazzei, Diego},
	title = {Vickybot, a Chatbot for Anxiety-Depressive Symptoms and Work-Related Burnout in Primary Care and Health Care Professionals: Development, Feasibility, and Potential Effectiveness Studies},
	year = {2023},
	journal = {Journal of Medical Internet Research},
	volume = {25},
	doi = {10.2196/43293},
	type = {Article},
	publication_stage = {Final},
	source = {Scopus},
	note = {Cited by: 19; All Open Access, Gold Open Access}
}

@CONFERENCE{Wang2020378,
	author = {Wang, Ruyi and Wang, Jiankun and Liao, Yuan and Wang, Jinyu},
	title = {Supervised machine learning chatbots for perinatal mental healthcare},
	year = {2020},
	journal = {Proceedings - 2020 International Conference on Intelligent Computing and Human-Computer Interaction, ICHCI 2020},
	pages = {378 – 383},
	doi = {10.1109/ICHCI51889.2020.00086},
	type = {Conference paper},
	publication_stage = {Final},
	source = {Scopus},
	note = {Cited by: 22}
}

@ARTICLE{Suharwardy2023,
	author = {Suharwardy, Sanaa and Ramachandran, Maya and Leonard, Stephanie A. and Gunaseelan, Anita and Lyell, Deirdre J. and Darcy, Alison and Robinson, Athena and Judy, Amy},
	title = {Feasibility and impact of a mental health chatbot on postpartum mental health: a randomized controlled trial},
	year = {2023},
	journal = {AJOG Global Reports},
	volume = {3},
	number = {3},
	doi = {10.1016/j.xagr.2023.100165},
	type = {Article},
	publication_stage = {Final},
	source = {Scopus},
	note = {Cited by: 10; All Open Access, Gold Open Access}
}

@ARTICLE{Pangsrisomboon202246,
	author = {Pangsrisomboon, Prima and Pyae, Aung and Thawitsri, Noppasorn and Liulak, Supasin},
	title = {Design and Development of an NLP-Based Mental Health Pre-screening Tool for Undergraduate Students in Thailand: A Usability Study},
	year = {2022},
	journal = {Communications in Computer and Information Science},
	volume = {1626 CCIS},
	pages = {46 – 60},
	doi = {10.1007/978-3-031-14832-3_4},
	type = {Conference paper},
	publication_stage = {Final},
	source = {Scopus},
	note = {Cited by: 0}
}

@ARTICLE{Fulmer2018,
	author = {Fulmer, Russell and Joerin, Angela and Gentile, Breanna and Lakerink, Lysanne and Rauws, Michiel},
	title = {Using psychological artificial intelligence (tess) to relieve symptoms of depression and anxiety: Randomized controlled trial},
	year = {2018},
	journal = {JMIR Mental Health},
	volume = {5},
	number = {4},
	doi = {10.2196/mental.9782},
	type = {Article},
	publication_stage = {Final},
	source = {Scopus},
	note = {Cited by: 284; All Open Access, Gold Open Access, Green Open Access}
}

@CONFERENCE{Kettle2021791,
	author = {Kettle, Liam and Lee, Yi-Ching},
	title = {“Welcome to your Daily Wellness Check”: The Proposed Evaluation of a SMS-based Conversational Agent for Managing Health and Wellbeing},
	year = {2021},
	journal = {Proceedings of the Human Factors and Ergonomics Society},
	volume = {65},
	number = {1},
	pages = {791 – 795},
	doi = {10.1177/1071181321651297},
	type = {Conference paper},
	publication_stage = {Final},
	source = {Scopus},
	note = {Cited by: 0; All Open Access, Bronze Open Access}
}

@ARTICLE{Dosovitsky2021,
	author = {Dosovitsky, Gilly and Bunge, Eduardo L.},
	title = {Bonding With Bot: User Feedback on a Chatbot for Social Isolation},
	year = {2021},
	journal = {Frontiers in Digital Health},
	volume = {3},
	doi = {10.3389/fdgth.2021.735053},
	type = {Article},
	publication_stage = {Final},
	source = {Scopus},
	note = {Cited by: 32; All Open Access, Gold Open Access, Green Open Access}
}

@ARTICLE{Pérez-Marín2013955,
	author = {Pérez-Marín, Diana and Pascual-Nieto, Ismael},
	title = {An exploratory study on how children interact with pedagogic conversational agents},
	year = {2013},
	journal = {Behaviour and Information Technology},
	volume = {32},
	number = {9},
	pages = {955 – 964},
	doi = {10.1080/0144929X.2012.687774},
	type = {Article},
	publication_stage = {Final},
	source = {Scopus},
	note = {Cited by: 28}
}

@CONFERENCE{Yang2019,
	author = {Yang, Xi and Aurisicchio, Marco and Baxter, Weston},
	title = {Understanding affective experiences with conversational agents},
	year = {2019},
	journal = {Conference on Human Factors in Computing Systems - Proceedings},
	doi = {10.1145/3290605.3300772},
	type = {Conference paper},
	publication_stage = {Final},
	source = {Scopus},
	note = {Cited by: 65}
}

@ARTICLE{Christoforakos2021,
	author = {Christoforakos, Lara and Feicht, Nina and Hinkofer, Simone and Löscher, Annalena and Schlegl, Sonja F. and Diefenbach, Sarah},
	title = {Connect With Me. Exploring Influencing Factors in a Human-Technology Relationship Based on Regular Chatbot Use},
	year = {2021},
	journal = {Frontiers in Digital Health},
	volume = {3},
	doi = {10.3389/fdgth.2021.689999},
	type = {Article},
	publication_stage = {Final},
	source = {Scopus},
	note = {Cited by: 9; All Open Access, Gold Open Access}
}

@CONFERENCE{Sayis2024945,
	author = {Sayis, Batuhan and Gunes, Hatice},
	title = {Technology-assisted Journal Writing for Improving Student Mental Wellbeing: Humanoid Robot vs. Voice Assistant},
	year = {2024},
	journal = {ACM/IEEE International Conference on Human-Robot Interaction},
	pages = {945 – 949},
	doi = {10.1145/3610978.3640721},
	type = {Conference paper},
	publication_stage = {Final},
	source = {Scopus},
	note = {Cited by: 0; All Open Access, Hybrid Gold Open Access}
}

@ARTICLE{Peuters2024,
	author = {Peuters, Carmen and Maenhout, Laura and Cardon, Greet and De Paepe, Annick and DeSmet, Ann and Lauwerier, Emelien and Leta, Kenji and Crombez, Geert},
	title = {A mobile healthy lifestyle intervention to promote mental health in adolescence: a mixed-methods evaluation},
	year = {2024},
	journal = {BMC Public Health},
	volume = {24},
	number = {1},
	doi = {10.1186/s12889-023-17260-9},
	type = {Article},
	publication_stage = {Final},
	source = {Scopus},
	note = {Cited by: 1; All Open Access, Gold Open Access}
}

@ARTICLE{Kuhlmeier202230,
	author = {Kuhlmeier, Florian Onur and Gnewuch, Ulrich and Lüttke, Stefan and Brakemeier, Eva-Lotta and Mädche, Alexander},
	title = {A Personalized Conversational Agent to Treat Depression in Youth and Young Adults – A Transdisciplinary Design Science Research Project},
	year = {2022},
	journal = {Lecture Notes in Computer Science (including subseries Lecture Notes in Artificial Intelligence and Lecture Notes in Bioinformatics)},
	volume = {13229 LNCS},
	pages = {30 – 41},
	doi = {10.1007/978-3-031-06516-3_3},
	type = {Conference paper},
	publication_stage = {Final},
	source = {Scopus},
	note = {Cited by: 2}
}

@ARTICLE{Hopman2023,
	author = {Hopman, Katherine and Richards, Deborah and Norberg, Melissa M.},
	title = {A Digital Coach to Promote Emotion Regulation Skills},
	year = {2023},
	journal = {Multimodal Technologies and Interaction},
	volume = {7},
	number = {6},
	doi = {10.3390/mti7060057},
	type = {Article},
	publication_stage = {Final},
	source = {Scopus},
	note = {Cited by: 3; All Open Access, Gold Open Access}
}

@ARTICLE{Darcy2021,
	author = {Darcy, Alison and Daniels, Jade and Salinger, David and Wicks, Paul and Robinson, Athena},
	title = {Evidence of human-level bonds established with a digital conversational agent: Cross-sectional, retrospective observational study},
	year = {2021},
	journal = {JMIR Formative Research},
	volume = {5},
	number = {5},
	doi = {10.2196/27868},
	type = {Article},
	publication_stage = {Final},
	source = {Scopus},
	note = {Cited by: 79; All Open Access, Gold Open Access}
}

@ARTICLE{Maharjan2022,
	author = {Maharjan, Raju and Doherty, Kevin and Rohani, Darius Adam and Bækgaard, Per and Bardram, Jakob E.},
	title = {Experiences of a Speech-enabled Conversational Agent for the Self-report of Well-being among People Living with Affective Disorders: An In-the-Wild Study},
	year = {2022},
	journal = {ACM Transactions on Interactive Intelligent Systems},
	volume = {12},
	number = {2},
	doi = {10.1145/3484508},
	type = {Article},
	publication_stage = {Final},
	source = {Scopus},
	note = {Cited by: 11; All Open Access, Bronze Open Access}
}

@CONFERENCE{Moilanen2022138,
	author = {Moilanen, Joonas and Visuri, Aku and Suryanarayana, Sharadhi Alape and Alorwu, Andy and Yatani, Koji and Hosio, Simo},
	title = {Measuring the Effect of Mental Health Chatbot Personality on User Engagement},
	year = {2022},
	journal = {ACM International Conference Proceeding Series},
	pages = {138 – 150},
	doi = {10.1145/3568444.3568464},
	type = {Conference paper},
	publication_stage = {Final},
	source = {Scopus},
	note = {Cited by: 6; All Open Access, Bronze Open Access}
}

@ARTICLE{Gabrielli2021,
	author = {Gabrielli, Silvia and Rizzi, Silvia and Bassi, Giulia and Carbone, Sara and Maimone, Rosa and Marchesoni, Michele and Forti, Stefano},
	title = {Engagement and effectiveness of a healthy-coping intervention via chatbot for university students during the COVID-19 pandemic: Mixed methods proof-of-concept study},
	year = {2021},
	journal = {JMIR mHealth and uHealth},
	volume = {9},
	number = {5},
	doi = {10.2196/27965},
	type = {Article},
	publication_stage = {Final},
	source = {Scopus},
	note = {Cited by: 54; All Open Access, Gold Open Access, Green Open Access}
}

@ARTICLE{Gkinko20221714,
	author = {Gkinko, Lorentsa and Elbanna, Amany},
	title = {Hope, tolerance and empathy: employees' emotions when using an AI-enabled chatbot in a digitalised workplace},
	year = {2022},
	journal = {Information Technology and People},
	volume = {35},
	number = {6},
	pages = {1714 – 1743},
	doi = {10.1108/ITP-04-2021-0328},
	type = {Article},
	publication_stage = {Final},
	source = {Scopus},
	note = {Cited by: 38; All Open Access, Hybrid Gold Open Access}
}

@ARTICLE{Ehrlich2024435,
	author = {Ehrlich, Christian and Hennelly, Sarah E. and Wilde, Natalie and Lennon, Oliver and Beck, Alan and Messenger, Hazel and Sergiou, Kat and Davies, Emma L.},
	title = {Evaluation of an Artificial Intelligence Enhanced Application for Student Wellbeing: Pilot Randomised Trial of the Mind Tutor},
	year = {2024},
	journal = {International Journal of Applied Positive Psychology},
	volume = {9},
	number = {1},
	pages = {435 – 454},
	doi = {10.1007/s41042-023-00133-2},
	type = {Article},
	publication_stage = {Final},
	source = {Scopus},
	note = {Cited by: 0; All Open Access, Hybrid Gold Open Access}
}

@ARTICLE{Terblanche202220,
	author = {Terblanche, Nicky and Molyn, Joanna and De Haan, Erik and Nilsson, Victor O.},
	title = {Coaching at Scale: Investigating theEfficacy of Artificial Intelligence Coaching},
	year = {2022},
	journal = {International Journal of Evidence Based Coaching and Mentoring},
	volume = {20},
	number = {2},
	pages = {20 – 36},
	doi = {10.24384/5cgf-ab69},
	type = {Article},
	publication_stage = {Final},
	source = {Scopus},
	note = {Cited by: 13}
}

@CONFERENCE{Lin2023264,
	author = {Lin, Shuya and Lin, Lingfeng and Hou, Cuiqin and Chen, Baijun and Li, Jianfeng and Ni, Shiguang},
	title = {Empathy-Based communication Framework for Chatbots: A Mental Health Chatbot Application and Evaluation},
	year = {2023},
	journal = {ACM International Conference Proceeding Series},
	pages = {264 – 272},
	doi = {10.1145/3623809.3623865},
	url = {https://www.scopus.com/inward/record.uri?eid=2-s2.0-85180128751&doi=10.1145%2f3623809.3623865&partnerID=40&md5=499acbb2f1af9735c163f0718c8ba86a},
	type = {Conference paper},
	publication_stage = {Final},
	source = {Scopus},
	note = {Cited by: 0; All Open Access, Bronze Open Access}
}

@ARTICLE{Suganuma2018,
	author = {Suganuma, Shinichiro and Sakamoto, Daisuke and Shimoyama, Haruhiko},
	title = {An embodied conversational agent for unguided internet-based cognitive behavior therapy in preventative mental health: Feasibility and acceptability pilot trial},
	year = {2018},
	journal = {JMIR Mental Health},
	volume = {5},
	number = {3},
	doi = {10.2196/10454},
	url = {https://www.scopus.com/inward/record.uri?eid=2-s2.0-85054279527&doi=10.2196%2f10454&partnerID=40&md5=8f1735446b600577cfac1c29a633abf6},
	type = {Article},
	publication_stage = {Final},
	source = {Scopus},
	note = {Cited by: 77; All Open Access, Gold Open Access, Green Open Access}
}

@CONFERENCE{Sia202134,
	author = {Sia, Dominic Ethan and Yu, Marco Jalen and Daliva, Justine Leo and Montenegro, Jaycee and Ong, Ethel},
	title = {Investigating the Acceptability and Perceived Effectiveness of a Chatbot in Helping Students Assess their Well-being},
	year = {2021},
	journal = {5th Asian CHI Symposium 2021},
	pages = {34 – 40},
	doi = {10.1145/3429360.3468177},
	url = {https://www.scopus.com/inward/record.uri?eid=2-s2.0-85114871694&doi=10.1145%2f3429360.3468177&partnerID=40&md5=0ae468f434c5737ff309b0518059c156},
	type = {Conference paper},
	publication_stage = {Final},
	source = {Scopus},
	note = {Cited by: 10}
}

@ARTICLE{Lee2023,
	author = {Lee, Minha and Contreras Alejandro, Jessica and IJsselsteijn, Wijnand},
	title = {Cultivating Gratitude with a Chatbot},
	year = {2023},
	journal = {International Journal of Human-Computer Interaction},
	doi = {10.1080/10447318.2023.2231277},
	url = {https://www.scopus.com/inward/record.uri?eid=2-s2.0-85165319216&doi=10.1080%2f10447318.2023.2231277&partnerID=40&md5=f91a0c0f9a52d056afbadce9870277ea},
	type = {Article},
	publication_stage = {Article in press},
	source = {Scopus},
	note = {Cited by: 1; All Open Access, Hybrid Gold Open Access}
}

@ARTICLE{Hungerbuehler2021,
	author = {Hungerbuehler, Ines and Daley, Kate and Cavanagh, Kate and Claro, Heloísa Garcia and Kapps, Michael},
	title = {Chatbot-based assessment of employees’ mental health: Design process and pilot implementation},
	year = {2021},
	journal = {JMIR Formative Research},
	volume = {5},
	number = {4},
	doi = {10.2196/21678},
	url = {https://www.scopus.com/inward/record.uri?eid=2-s2.0-85104845527&doi=10.2196%2f21678&partnerID=40&md5=2f5fb9ca41edc5f84cd4708f2f9fdf21},
	type = {Article},
	publication_stage = {Final},
	source = {Scopus},
	note = {Cited by: 36; All Open Access, Gold Open Access, Green Open Access}
}

@ARTICLE{Inkster2023,
	author = {Inkster, Becky and Kadaba, Madhura and Subramanian, Vinod},
	title = {Understanding the impact of an AI-enabled conversational agent mobile app on users’ mental health and wellbeing with a self-reported maternal event: a mixed method real-world data mHealth study},
	year = {2023},
	journal = {Frontiers in Global Women's Health},
	volume = {4},
	doi = {10.3389/fgwh.2023.1084302},
	url = {https://www.scopus.com/inward/record.uri?eid=2-s2.0-85161981605&doi=10.3389%2ffgwh.2023.1084302&partnerID=40&md5=634e39e693d426e16d714541146688e6},
	type = {Article},
	publication_stage = {Final},
	source = {Scopus},
	note = {Cited by: 3; All Open Access, Gold Open Access, Green Open Access}
}

@ARTICLE{Lopatovska2022192,
	author = {Lopatovska, Irene and Turpin, Olivia and Yoon, Ji Hee and Brown, Diedre and Vroom, Laura and Nielsen, Craig and Hayes, Kelli and Roslund, Karin and Dickson, Mary and Anger, Daniel},
	title = {Measuring the Impact of Conversational Technology Interventions on Adolescent Wellbeing: Quantitative and Qualitative Approaches},
	year = {2022},
	journal = {Proceedings of the Association for Information Science and Technology},
	volume = {59},
	number = {1},
	pages = {192 – 204},
	doi = {10.1002/pra2.639},
	url = {https://www.scopus.com/inward/record.uri?eid=2-s2.0-85139955983&doi=10.1002%2fpra2.639&partnerID=40&md5=59d8aba47e9396962e3c24a588f4b7d1},
	type = {Article},
	publication_stage = {Final},
	source = {Scopus},
	note = {Cited by: 0}
}

@ARTICLE{Shidara2024,
	author = {Shidara, Kazuhiro and Tanaka, Hiroki and Adachi, Hiroyoshi and Kanayama, Daisuke and Kudo, Takashi and Nakamura, Satoshi},
	title = {Adapting the Number of Questions Based on Detected Psychological Distress for Cognitive Behavioral Therapy With an Embodied Conversational Agent: Comparative Study},
	year = {2024},
	journal = {JMIR Formative Research},
	volume = {8},
	doi = {10.2196/50056},
	url = {https://www.scopus.com/inward/record.uri?eid=2-s2.0-85191202268&doi=10.2196%2f50056&partnerID=40&md5=f49f79ee7cd8a553ac415614797c4675},
	type = {Article},
	publication_stage = {Final},
	source = {Scopus},
	note = {Cited by: 0; All Open Access, Gold Open Access}
}

@ARTICLE{Durden2023,
	author = {Durden, Emily and Pirner, Maddison C. and Rapoport, Stephanie J. and Williams, Andre and Robinson, Athena and Forman-Hoffman, Valerie L.},
	title = {Changes in stress, burnout, and resilience associated with an 8-week intervention with relational agent “Woebot”},
	year = {2023},
	journal = {Internet Interventions},
	volume = {33},
	doi = {10.1016/j.invent.2023.100637},
	url = {https://www.scopus.com/inward/record.uri?eid=2-s2.0-85162147570&doi=10.1016%2fj.invent.2023.100637&partnerID=40&md5=10f60925c0f6c5fdc6c3bd59a93f9cd9},
	type = {Article},
	publication_stage = {Final},
	source = {Scopus},
	note = {Cited by: 13; All Open Access, Gold Open Access}
}

@ARTICLE{Daley2020,

AUTHOR={Daley, Kate  and Hungerbuehler, Ines  and Cavanagh, Kate  and Claro, Heloísa Garcia  and Swinton, Paul Alan  and Kapps, Michael },

TITLE={Preliminary Evaluation of the Engagement and Effectiveness of a Mental Health Chatbot},

JOURNAL={Frontiers in Digital Health},

VOLUME={2},

YEAR={2020},

URL={https://www.frontiersin.org/journals/digital-health/articles/10.3389/fdgth.2020.576361},

DOI={10.3389/fdgth.2020.576361},

ISSN={2673-253X}}

@article{Allan2023,
author = {D. D. Allan},
title = {A mobile messaging-based conversational agent-led stress mindset intervention for New Zealand small-to-medium-sized enterprise owner-managers: effectiveness and acceptability study},
journal = {Behaviour \& Information Technology},
volume = {0},
number = {0},
pages = {1--14},
year = {2023},
publisher = {Taylor \& Francis},
doi = {10.1080/0144929X.2023.2259500},


URL = { 
    
        https://doi.org/10.1080/0144929X.2023.2259500
    
    

},
eprint = { 
    
        https://doi.org/10.1080/0144929X.2023.2259500
    
    

}

}

@article{Eagle2022,
author = {Eagle, Tessa and Blau, Conrad and Bales, Sophie and Desai, Noopur and Li, Victor and Whittaker, Steve},
title = {“I don’t know what you mean by `I am anxious'”: A New Method for Evaluating Conversational Agent Responses to Standardized Mental Health Inputs for Anxiety and Depression},
year = {2022},
issue_date = {June 2022},
publisher = {Association for Computing Machinery},
address = {New York, NY, USA},
volume = {12},
number = {2},
issn = {2160-6455},
url = {https://doi.org/10.1145/3488057},
doi = {10.1145/3488057},
journal = {ACM Trans. Interact. Intell. Syst.},
month = {jul},
articleno = {12},
numpages = {23}
}

@inproceedings{Romanovskyi2021ElomiaCT,
  title={Elomia Chatbot: The Effectiveness of Artificial Intelligence in the Fight for Mental Health},
  author={Oleksandr Romanovskyi and Nina Pidbutska and Anastasiia Knysh},
  booktitle={International Conference on Computational Linguistics and Intelligent Systems},
  year={2021},
  url={https://api.semanticscholar.org/CorpusID:235271819}
}

@inproceedings{Schroeder2018,
author = {Schroeder, Jessica and Wilkes, Chelsey and Rowan, Kael and Toledo, Arturo and Paradiso, Ann and Czerwinski, Mary and Mark, Gloria and Linehan, Marsha M.},
title = {Pocket Skills: A Conversational Mobile Web App To Support Dialectical Behavioral Therapy},
year = {2018},
isbn = {9781450356206},
publisher = {Association for Computing Machinery},
address = {New York, NY, USA},
url = {https://doi.org/10.1145/3173574.3173972},
doi = {10.1145/3173574.3173972},
booktitle = {Proceedings of the 2018 CHI Conference on Human Factors in Computing Systems},
pages = {1–15},
numpages = {15},
location = {Montreal QC, Canada},
series = {CHI '18}
}

@InProceedings{Oliveira2021,
author="Oliveira, Alana Lucia Souza
and Matos, Leonardo Nogueira
and Junior, Methanias Cola{\c{c}}o
and Delabrida, Zenith Nara Costa",
editor="Gervasi, Osvaldo
and Murgante, Beniamino
and Misra, Sanjay
and Garau, Chiara
and Ble{\v{c}}i{\'{c}}, Ivan
and Taniar, David
and Apduhan, Bernady O.
and Rocha, Ana Maria A. C.
and Tarantino, Eufemia
and Torre, Carmelo Maria",
title="An Initial Assessment of a Chatbot for Rumination-Focused Cognitive Behavioral Therapy (RFCBT) in College Students",
booktitle="Computational Science and Its Applications -- ICCSA 2021",
year="2021",
publisher="Springer International Publishing",
address="Cham",
pages="549--564",
isbn="978-3-030-86979-3"
}

@article{Mauriello2021,
Author = {Mauriello, Matthew Louis and Tantivasadakarn, Nantanick and
   Mora-Mendoza, Marco Antonio and Lincoln, Emmanuel Thierry and Hon, Grace
   and Nowruzi, Parsa and Simon, Dorien and Hansen, Luke and Goenawan,
   Nathaniel H. and Kim, Joshua and Gowda, Nikhil and Jurafsky, Dan and
   Paredes, Pablo Enrique},
Title = {A Suite of Mobile Conversational Agents for Daily Stress Management
   (Popbots): Mixed Methods Exploratory Study},
Journal = {JMIR FORMATIVE RESEARCH},
Year = {2021},
Volume = {5},
Number = {9},
Month = {SEP},
DOI = {10.2196/25294},
Article-Number = {e25294},
EISSN = {2561-326X},
ORCID-Numbers = {Paredes, Pablo/0000-0003-2431-9190
   Nowruzi, Parsa/0000-0001-5372-5513
   Mauriello, Matthew Louis/0000-0001-5359-6520
   Hon, Grace/0000-0001-9471-2209
   Tantivasadakarn, Nantanick/0000-0002-9895-355X
   Mora-Mendoza, Marco/0000-0002-7305-3520},
Unique-ID = {WOS:000853674000006},
}

@article{Schick2022,
Author = {Schick, Anita and Feine, Jasper and Morana, Stefan and Maedche,
   Alexander and Reininghaus, Ulrich},
Title = {Validity of Chatbot Use for Mental Health Assessment: Experimental Study},
Journal = {JMIR MHEALTH AND UHEALTH},
Year = {2022},
Volume = {10},
Number = {10},
Month = {OCT},
DOI = {10.2196/28082},
Article-Number = {e28082},
ISSN = {2291-5222},
ResearcherID-Numbers = {Morana, Stefan/AAL-6118-2020
   Reininghaus, Ulrich/ABE-1130-2020
   },
ORCID-Numbers = {Morana, Stefan/0000-0002-2266-1286
   Reininghaus, Ulrich/0000-0002-9227-5436
   Schick, Anita/0000-0002-2043-0353
   Maedche, Alexander/0000-0001-6546-4816},
Unique-ID = {WOS:001126337300001},
}

@ARTICLE{Karkosz2022424,
	author = {Karkosz, Stanisław and Michałowski, Jarosław M. and Sanna, Katarzyna and Szczepaniak, Norbert and Konat, Barbara},
	title = {Human-Therapeutic Chatbot Interaction Analysis, on the Example of Fido},
	year = {2022},
	journal = {Communications in Computer and Information Science},
	volume = {1583 CCIS},
	pages = {424 – 429},
	doi = {10.1007/978-3-031-06394-7_53},
	url = {https://www.scopus.com/inward/record.uri?eid=2-s2.0-85133174351&doi=10.1007%2f978-3-031-06394-7_53&partnerID=40&md5=66d78f75d8d3a908a41f87a33204d5b1},
	type = {Conference paper},
	publication_stage = {Final},
	source = {Scopus},
	note = {Cited by: 1}
}

@CONFERENCE{Ghandeharioun20198,
	author = {Ghandeharioun, Asma and McDuff, Daniel and Czerwinski, Mary and Rowan, Kael},
	title = {Towards Understanding Emotional Intelligence for Behavior Change Chatbots},
	year = {2019},
	journal = {2019 8th International Conference on Affective Computing and Intelligent Interaction, ACII 2019},
	pages = {8 – 14},
	doi = {10.1109/ACII.2019.8925433},
	url = {https://www.scopus.com/inward/record.uri?eid=2-s2.0-85077801754&doi=10.1109%2fACII.2019.8925433&partnerID=40&md5=e758d5ed4d6a13c3ad8ca39f166531e0},
	type = {Conference paper},
	publication_stage = {Final},
	source = {Scopus},
	note = {Cited by: 39; All Open Access, Green Open Access}
}

@CONFERENCE{Kurashige2019962,
	author = {Kurashige, Kentarou and Tsuruta, Setsuo and Sakurai, Eriko and Sakurai, Yoshitaka and Knauf, Rainer and Damiani, Ernesto and Kutics, Andrea},
	title = {Robotized Counselor Evaluation using Linguistic Detection of Feeling Polarity Change},
	year = {2019},
	journal = {2019 IEEE Symposium Series on Computational Intelligence, SSCI 2019},
	pages = {962 – 967},
	doi = {10.1109/SSCI44817.2019.9002842},
	url = {https://www.scopus.com/inward/record.uri?eid=2-s2.0-85080857328&doi=10.1109%2fSSCI44817.2019.9002842&partnerID=40&md5=489ed7b23b26a8a9c513b93762176d57},
	type = {Conference paper},
	publication_stage = {Final},
	source = {Scopus},
	note = {Cited by: 0}
}

@ARTICLE{Ta2020,
	author = {Ta, Vivian and Griffith, Caroline and Boatfield, Carolynn and Wang, Xinyu and Civitello, Maria and Bader, Haley and DeCero, Esther and Loggarakis, Alexia},
	title = {User experiences of social support from companion chatbots in everyday contexts: Thematic analysis},
	year = {2020},
	journal = {Journal of Medical Internet Research},
	volume = {22},
	number = {3},
	doi = {10.2196/16235},
	url = {https://www.scopus.com/inward/record.uri?eid=2-s2.0-85081529648&doi=10.2196%2f16235&partnerID=40&md5=0665361febb5e0b2f1d4266e29019f34},
	type = {Article},
	publication_stage = {Final},
	source = {Scopus},
	note = {Cited by: 136; All Open Access, Gold Open Access, Green Open Access}
}

@ARTICLE{Ciechanowski2019539,
	author = {Ciechanowski, Leon and Przegalinska, Aleksandra and Magnuski, Mikolaj and Gloor, Peter},
	title = {In the shades of the uncanny valley: An experimental study of human–chatbot interaction},
	year = {2019},
	journal = {Future Generation Computer Systems},
	volume = {92},
	pages = {539 – 548},
	doi = {10.1016/j.future.2018.01.055},
	url = {https://www.scopus.com/inward/record.uri?eid=2-s2.0-85042030188&doi=10.1016%2fj.future.2018.01.055&partnerID=40&md5=96b4e304f9ac65c19548550b98bd979c},
	type = {Article},
	publication_stage = {Final},
	source = {Scopus},
	note = {Cited by: 339}
}

@inproceedings{Feng2023,
Publication-Type = {J},
Type = {Conference Paper},
Title = {PCDialogEval: Persona and Context Aware Emotional Dialogue Evaluation},
Author = {Feng, Y. and Wang, L. and Cao, Z. and He, L.},
Editor = {Iliadis, L. and Papaleonidas, A. and Angelov, P. and Jayne, C.},
Booktitle = {Artificial Neural Networks and Machine Learning - ICANN 2023: 32nd
   International Conference on Artificial Neural Networks, Proceedings.
   Lecture Notes in Computer Science (14262)},
Month = {2023},
Year = {2023},
Pages = {152-65},
Note = {International Conference on Artificial Neural Networks, 26-29 Sept.
   2023, Heraklion, Greece},
ISBN = {978-3-031-44201-8},
Identifying-Codes = {[10.1007/978-3-031-44201-8\_13]},
Unique-ID = {INSPEC:24327847},
}

@ARTICLE{Jabir2023,
	author = {Jabir, Ahmad Ishqi and Martinengo, Laura and Lin, Xiaowen and Torous, John and Subramaniam, Mythily and Car, Lorainne Tudor},
	title = {Evaluating Conversational Agents for Mental Health: Scoping Review of Outcomes and Outcome Measurement Instruments},
	year = {2023},
	journal = {Journal of Medical Internet Research},
	volume = {25},
	doi = {10.2196/44548},
	url = {https://www.scopus.com/inward/record.uri?eid=2-s2.0-85153121663&doi=10.2196%2f44548&partnerID=40&md5=125289c5ebabfb90160dcd459cbd8fd7},
	type = {Review},
	publication_stage = {Final},
	source = {Scopus},
	note = {Cited by: 10; All Open Access, Gold Open Access, Green Open Access}
}

@ARTICLE{Khosravi2024,
	author = {Khosravi, Mohsen and Azar, Ghazaleh},
	title = {Factors influencing patient engagement in mental health chatbots: A thematic analysis of findings from a systematic review of reviews},
	year = {2024},
	journal = {Digital Health},
	volume = {10},
	doi = {10.1177/20552076241247983},
	url = {https://www.scopus.com/inward/record.uri?eid=2-s2.0-85191444647&doi=10.1177%2f20552076241247983&partnerID=40&md5=e892334dd62a7ecbc45e3d13376d3a1a},
	type = {Review},
	publication_stage = {Final},
	source = {Scopus},
	note = {Cited by: 0; All Open Access, Gold Open Access}
}

@Article{Kocaballi2019,
author="Kocaballi, Ahmet Baki
and Berkovsky, Shlomo
and Quiroz, Juan C
and Laranjo, Liliana
and Tong, Huong Ly
and Rezazadegan, Dana
and Briatore, Agustina
and Coiera, Enrico",
title="The Personalization of Conversational Agents in Health Care: Systematic Review",
journal="J Med Internet Res",
year="2019",
month="Nov",
day="7",
volume="21",
number="11",
pages="e15360",
issn="1438-8871",
doi="10.2196/15360",
url="https://www.jmir.org/2019/11/e15360",
url="https://doi.org/10.2196/15360",
url="http://www.ncbi.nlm.nih.gov/pubmed/31697237"
}

@article{Ya-Hsin2023,
author = {Chou, Ya-Hsin and Lin, Chemin and Lee, Shwu-Hua and Chien, Ya-Wen Chang and Cheng, Li-Chen},
title = {Potential Mobile Health Applications for Improving the Mental Health of the Elderly: A Systematic Review},
journal = {Clinical Interventions in Aging},
volume = {18},
number = {},
pages = {1523--1534},
year = {2023},
publisher = {Dove Medical Press},
doi = {10.2147/CIA.S410396},

    note ={PMID: 37727447},


URL = { 
    
    
        https://www.tandfonline.com/doi/abs/10.2147/CIA.S410396
    

},
eprint = { 
    
    
        https://www.tandfonline.com/doi/pdf/10.2147/CIA.S410396
    

}

}

@ARTICLE{Lin2023,
	author = {Lin, Xiaowen and Martinengo, Laura and Jabir, Ahmad Ishqi and Ho, Andy Hau Yan and Car, Josip and Atun, Rifat and Car, Lorainne Tudor},
	title = {Scope, Characteristics, Behavior Change Techniques, and Quality of Conversational Agents for Mental Health and Well-Being: Systematic Assessment of Apps},
	year = {2023},
	journal = {Journal of Medical Internet Research},
	volume = {25},
	doi = {10.2196/45984},
	url = {https://www.scopus.com/inward/record.uri?eid=2-s2.0-85165518096&doi=10.2196%2f45984&partnerID=40&md5=60f5eabebd75bf8f1148c488f0e01698},
	type = {Article},
	publication_stage = {Final},
	source = {Scopus},
	note = {Cited by: 3; All Open Access, Gold Open Access, Green Open Access}
}

@ARTICLE{Haque2023,
	author = {Haque, M.D. Romael and Rubya, Sabirat},
	title = {An Overview of Chatbot-Based Mobile Mental Health Apps: Insights From App Description and User Reviews},
	year = {2023},
	journal = {JMIR mHealth and uHealth},
	volume = {11},
	doi = {10.2196/44838},
	url = {https://www.scopus.com/inward/record.uri?eid=2-s2.0-85159826255&doi=10.2196%2f44838&partnerID=40&md5=0ff14efad8231f46d5ec6ced3aecc305},
	type = {Article},
	publication_stage = {Final},
	source = {Scopus},
	note = {Cited by: 38; All Open Access, Gold Open Access, Green Open Access}
}

@phdthesis{Liao2021,
  author  = {Liao, Yuting
    and St.~Jean, Beth
    and Lazar, Amanda
    and Choe, Kyoung Eun
    and Kivlighan, Dennis},
  advisor = {Vitak, Jessica},
  title   = {Design and Evaluation of a Conversational Agent for Mental Health Support: Forming Human-Agent Sociotechnical and Therapeutic Relationships},
  year    = {2021},
  school  = {University of Maryland, College Park},
  note    = {AAI28717215},
  isbn    = {9798790625947}
}

@ARTICLE{Franco2023,
author = {Franco D'Souza, Russell and Amanullah, Shabbir and Mathew, Mary and Surapaneni, Krishna Mohan},
title = {Appraising the performance of ChatGPT in psychiatry using 100 clinical case vignettes},
year = {2023},
journal = {Asian Journal of Psychiatry},
volume = {89},
doi = {10.1016/j.ajp.2023.103770},
url = {https://www.scopus.com/inward/record.uri?eid=2-s2.0-85173142546&doi=10.1016%2fj.ajp.2023.103770&partnerID=40&md5=c68727ba575fa35a55cccfc797f81c8b},
type = {Article},
	publication_stage = {Final},
source = {Scopus},
note = {Cited by: 13}
}

@misc{1904.09675,
Author = {Tianyi Zhang and Varsha Kishore and Felix Wu and Kilian Q. Weinberger and Yoav Artzi},
Title = {BERTScore: Evaluating Text Generation with BERT},
Year = {2019},
Eprint = {arXiv:1904.09675},
}

\appendix

\section{Search Strings}
\subsection{ACM Digital Library}
\textit{("mental health" OR "mental wellness" OR wellbeing OR "well-being" OR "SWB" OR happiness OR happy OR "positive affect*" OR "negative affect*" OR "positive emotion*" OR "negative emotion*" OR mood OR "life satisfaction" OR "satisfaction with life") AND ("social bot*" OR "dialogue system*" OR "conversational agent*" OR "conversational bot*" OR "conversational system*" OR "conversational interface*" OR "chatbot*" OR "chat bot*" OR "chatterbot*" OR "chatter bot*" OR "chat-bot*" OR "smartbot*" OR "smart bot*" OR "smart-bot*" OR "digital assistant*" OR "counseling agent*" ) AND ("evaluation" OR "scale" OR "questionnaire" OR "guideline" OR "Review" OR "qualit*") AND NOT("autis*" OR "Alzheimer" OR "Parkinson" OR "disease" OR "*ache" OR "ADHD")}
\subsection{Scopus}
\textit{TITLE(("mental health" OR "mental wellness" OR wellbeing OR "well-being" OR "SWB" OR happiness OR happy OR "positive affect*" OR "negative affect*" OR "positive emotion*" OR "negative emotion*" OR mood OR "life satisfaction" OR "satisfaction with life") AND ("social bot*" OR "dialogue system*" OR "conversational agent*" OR "conversational bot*" OR "conversational system*" OR "conversational interface*" OR "chatbot*" OR "chat bot*" OR "chatterbot*" OR "chatter bot*" OR "chat-bot*" OR "smartbot*" OR "smart bot*" OR "smart-bot*" OR "digital assistant*" OR "counseling agent*" ) AND ("evaluation" OR "scale" OR "questionnaire" OR "guideline" OR "Review" OR "qualit*")AND NOT("autis*" OR "Alzheimer" OR "Parkinson" OR "disease" OR "*ache" OR "ADHD")) OR ABS(("mental health" OR "mental wellness" OR wellbeing OR "well-being" OR "SWB" OR happiness OR happy OR "positive affect*" OR "negative affect*" OR "positive emotion*" OR "negative emotion*" OR mood OR "life satisfaction" OR "satisfaction with life") AND ("social bot*" OR "dialogue system*" OR "conversational agent*" OR "conversational bot*" OR "conversational system*" OR "conversational interface*" OR "chatbot*" OR "chat bot*" OR "chatterbot*" OR "chatter bot*" OR "chat-bot*" OR "smartbot*" OR "smart bot*" OR "smart-bot*" OR "digital assistant*" OR "counseling agent*" ) AND ("evaluation" OR "scale" OR "questionnaire"OR "guideline" OR "Review" OR "qualit*")AND NOT("autis*" OR "Alzheimer" OR "Parkinson" OR "disease" OR "*ache" OR "ADHD"))}
\subsection{PsychInfo}
\textit{("mental health" OR "mental wellness" OR wellbeing OR "well-being" OR "SWB" OR happiness OR happy OR "positive affect*" OR "negative affect*" OR "positive emotion*" OR "negative emotion*" OR mood OR "life satisfaction" OR "satisfaction with life") AND ("social bot*" OR "dialogue system*" OR "conversational agent*" OR "conversational bot*" OR "conversational system*" OR "conversational interface*" OR "chatbot*" OR "chat bot*" OR "chatterbot*" OR "chatter bot*" OR "chat-bot*" OR "smartbot*" OR "smart bot*" OR "smart-bot*" OR "digital assistant*" OR "counseling agent*" ) AND ("evaluation" OR "scale" OR "questionnaire" OR "guideline" OR "Review" OR "qualit*") AND NOT("autis*" OR "Alzheimer" OR "Parkinson" OR "disease" OR "*ache" OR "ADHD")}

\end{document}